\documentclass[12pt]{article}

\usepackage{setspace,graphicx,epstopdf,amsmath,amsfonts,amssymb,amsthm}
\usepackage{marginnote,datetime,enumitem,subfigure,rotating,fancyvrb}
\usepackage{float}
\usepackage[
  colorlinks=true,
  linkcolor=blue,
  citecolor=blue]{hyperref}
\usepackage{booktabs}
\usepackage[longnamesfirst]{natbib}
\usepackage{mathtools}
\usepackage{lscape}
\usepackage[toc,page]{appendix}

\usepackage{tikz}
\usetikzlibrary{positioning}
\usepackage{forest}

\usetikzlibrary{trees}

\usdate

\hypersetup{draft=true}

\usepackage[margin=1in]{geometry}%

\newlength\mytemplength
\newcommand\parboxc[3]{%
    \settowidth{\mytemplength}{#3}%
    \parbox[#1][#2]{\mytemplength}{\centering #3}%
}

\makeatletter\let\chapter\@undefined\makeatother 

\usepackage{bbm}

\usepackage{indentfirst} 
\usepackage[labelfont=bf,labelsep=period]{caption}   
\def\QQ{{\mathbb Q}}

\usepackage{caption}

\usepackage{changes}
\definechangesauthor[name={HC}, color=red]{HC}

\begin{document}

\setlist{noitemsep}  
\onehalfspacing      
\parskip 3pt


\title{Deep Structural Estimation: \\ With an Application to Option Pricing}


\author{Hui Chen\thanks{Department of Finance, MIT Sloan School of Management and NBER. Email: huichen@mit.edu.} \and Antoine Didisheim\thanks{Department of Finance, HEC Lausanne, University of Lausanne and Swiss Finance Institute. Email: antoine.didisheim@unil.} \and Simon Scheidegger\thanks{Department of Economics, HEC Lausanne, University of Lausanne. Email: simon.scheidegger@unil.} \thanks{This work is generously supported by grants from the Swiss Platform for Advanced Scientific Computing (PASC) under project ID ``Computing equilibria in heterogeneous agent macro models on contemporary HPC platforms", the Swiss National Supercomputing Center (CSCS) under project ID 995, and the Swiss National Science Foundation under project IDs ``Can Economic Policy Mitigate Climate-Change?", and ``New methods for asset pricing with frictions". Simon Scheidegger gratefully acknowledges support from the MIT Sloan School of Management and the Cowles Foundation at Yale University.}}

\date{February 9, 2021}              


\maketitle
\thispagestyle{empty}

\bigskip

\begin{abstract}




We propose a novel structural estimation framework in which we train a surrogate of an economic model with deep neural networks. Our methodology alleviates the curse of dimensionality and speeds up the evaluation and parameter estimation by orders of magnitudes, which significantly enhances one's ability to conduct analyses that require frequent parameter re-estimation. As an empirical application, we compare two popular option pricing models (the Heston and the Bates model with double-exponential jumps) against a non-parametric random forest model. We document that: a) the Bates model produces better out-of-sample pricing on average, but both structural models fail to outperform random forest for large areas of the volatility surface; b) random forest is more competitive at short horizons (e.g., 1-day), for short-dated options (with less than 7 days to maturity), and on days with poor liquidity; c) both structural models outperform random forest in out-of-sample delta hedging; d) the Heston model's relative performance has deteriorated significantly after the 2008 financial crisis.

\end{abstract}

\medskip


\clearpage

\setcounter{page}{1}

\section{Introduction}
\label{sec:Intro}

Driven by theoretical developments, the availability of ``big data,'' and gains in computing power, contemporary models in economics and finance have seen tremendous growth in complexity. However, the state-of-the-art structural models often impose a substantial roadblock to researchers and practitioners: with an ever-increasing number of states and parameters, they increasingly suffer from the curse of dimensionality~\citep{Bellman1961AdaptiveTour}---that is, the computational burden to evaluate the model or estimate model parameters and hidden states grows exponentially with every additional degree of freedom.



Consequently, economists are often forced to sacrifice certain features of the model in order to reduce model dimensionality, estimate only a partial set of parameters while pre-fixing the others, and estimate the model only once using the full sample of data. Such restrictions limit a researcher's ability to carry out important model analysis, including: a) sub-sample or out-of-sample analyses, b) cross validation, in particular with time-series forecasting models, where it is routine to re-estimate the model using a moving window or expanding window to take into account the latest available data while avoiding any look-ahead bias \citep[see, e.g.,][]{10.1093/rfs/hhm014}, c) management of heterogeneity in data through repeated re-estimation, for example, when fitting a consumption-portfolio model to a large cross-section of households, d) testing for parameter stability. In some cases, the exponentially increasing computational costs can prevent the adoption of complex models by practitioners in real-time. 


To tackle these issues, we introduce \textit{deep structural estimation}, a new framework to more efficiently evaluate and estimate structural models. At its core, our method re-purposes the concept of surrogate models commonly applied in physics and engineering in the context of financial models.\footnote{In physics and engineering, Gaussian processes regression \citep[see, e.g.,][]{williams2006gaussian,tripathy2016gaussian,bilionis2012multi,bilionis2013multi,chen2015uncertainty}, radial basis functions~\citep{park1991universal}, or relevance vector machines~\citep{bilionis2012multidimensional} are often used to build surrogate models. More recently, following the rapid developments in the theory of stochastic optimization and artificial intelligence as well as the advances in computer hardware leading to the widespread availability of graphic processing units (GPUs; see, e.g.,~\cite{IPDPS2018,aldrich2011}, and references therein), researchers have turned their attention towards \textit{deep neural networks}~\citep[see, e.g.,][]{tripathy2018deep,liu2019neural}.} We adopt deep neural networks to create cheap-to-evaluate surrogates of high-dimensional models---that is, a function that takes the same input as the original model and yields the same output at a significantly lower computational cost. To construct these surrogates, we treat parameters as pseudo-states and estimate both parameters and hidden states from the data in a comparatively cheap optimization procedure. This methodology can alleviate the curse of dimensionality and reduce the computational cost of evaluating the model by orders of magnitude.

To see the intuition for the curse of dimensionality and how deep learning can help solve this problem, consider the task of approximating a univariate function by visiting $10$ locations along the input state (this is our training sample). For a bivariate function of comparable properties, one would need to visit about $10^2$ points to maintain a similar accuracy level. Generalizing to $d$ states, this procedure requires visiting $\mathcal{O}(10^d)$ locations in the $d$-dimensional state space and to evaluate the function at all those locations. Even in a situation where a single function evaluation is relatively inexpensive to compute, attempting naively to approximate a high-dimensional function in this way can quickly become infeasible. 

In contrast, the construction of a surrogate model based on neural networks alleviates the curse of dimensionality because one can typically train a deep network to accurately approximates the high-dimensional function we are concerned with using substantially fewer observations than one would use with more commonly used methods based on Cartesian grids. \cite{doi:10.1137/18M1189336}, for instance, show formally under certain assumptions that approximating multivariate functions by deep ReLU networks can be bounded by sparse grids (see, e.g.,~\cite{Bungartz2004SparseGrids}, and references therein), which themselves alleviate the curse of dimensionality. 

In one of the examples we study below, a double-exponential jump-diffusion model has 14 states ($d=14$). We train a surrogate using a DNN with seven hidden layers and 400 neurons in each layer, which correspond to nearly $10^6$ trainable parameters. The model is then successfully trained on a sample of size $N = 10^{9}$, far smaller than $10^{14}$. Our numerical experiments show that while it takes over 40 minutes to estimate this model using traditional grid-based methods, the surrogate model reduces the execution time to less than one second. Although comparing the time to execute an algorithm can be misleading (it depends on expert knowledge and code optimization), the gulf between the two methods in our experiments demonstrates the potential for deep structural estimation.

Such computational gains have important consequences for research. They allow for i) re-estimating a model's parameters and hidden states at a high frequency, ii) thorough out-of-sample analysis of model performance, which ensures the generalization of the results, and iii) robust statistical testing of parameters' stability.

As an application, we use the \textit{deep structural estimation} methodology to compare two approaches commonly used for pricing options: structural versus entirely data-driven. Conceptually, structural models have the advantage of linking the dynamics of option prices with the dynamics of the hidden state variables and the structural parameters. In contrast, while (reduced-form) data-driven models can be more flexible in fitting the data in-sample, they are more prone to over-fitting. Moreover, their respective performance relative to structural models could also deteriorate out-of-sample when a change in the underlying hidden state significantly alters the shape of the volatility surface; changes that a suitable structural model might be able to predict. 

For reasons of transparency, we select as examples of structural models the classic stochastic volatility model proposed by~\cite{heston1993closed} (HM), and a double-exponential jump-diffusion model extended from \cite{bates1996jumps} (BDJM). Both models have been extensively studied in the literature and are well-understood. They feature one hidden state variable: spot volatility. As an example of a data-driven model, we use a non-parametric \textit{practitioner's Black-Scholes} model by modeling the implied volatility surface using random forests~\citep[RFs; see, e.g.,][]{breiman:01}.


We create a surrogate for each of these option-pricing models. The surrogates enable us to re-estimate each option model's parameters as well as the spot volatility on the daily cross-section of S\&P500 options for every individual day for the last seventeen years to produce a time series of jointly estimated optimal parameters and hidden states across the sample. We stress that without the surrogate technology, the computational cost of estimating the parameters would render this analysis infeasible. 

Our empirical analysis reveals several findings. \textit{First}, we compare the in- and out-of-sample performance of the structural models and the RF model. As expected, each pricing model's in-sample performance is directly related to the number of degrees of freedom, with RF producing the smallest mean squared error (MSE) of the three option pricing models. Out of sample, BDJM outperforms HM and RF on average; RF outperforms HM at short horizons but underperforms when the forecasting horizon exceeds 7 days. In the cross section, RF outperforms both parametric models across all moneyness for options with a time to maturity less than one week. Furthermore, BDJM outperforms the other models when market volatility and jump risks are elevated, while the structural models perform worse for options with poor liquidity (as measured by bid-ask spreads). 
%
%
These results highlight the specific area of the volatility surface on which option pricing theory has room to improve.

\textit{Second}, using the statistical test proposed by \cite{andersen2015parametric}, we show that the estimated parameters for both structural models are highly unstable: the parameter estimates from two consecutive trading days differ significantly (at the 1\% level) 41.6\% of the time for the BDJM and 60.7\% of the time for the HM. Furthermore, the percentage of days with significant changes in their parameters increase with time. Note that the test was originally designed to compare two large samples, but the  \textit{deep surrogate estimation} framework allows us to apply it at a daily frequency. 

\textit{Third}, we compare the models' hedging performances. Indeed, one of the important tasks of option pricing models is to estimate the option delta, an option's price sensitivity to changes in the price of the underlying asset. This key estimation allows market makers and liquidity providers to hedge their position dynamically. For each option, we perform delta hedging at daily frequency based on the model-specific theoretical deltas calculated using parameter values estimated that day.  
%
%
We show that the BDJM replicating portfolio outperforms the HM on average. This difference is concentrated in short-dated out-of-money puts and long-dated out-of-money calls, and it mostly occurred after 2011. 
We also show that RFs are ill-suited for constructing hedging portfolios due to its non-smoothness. This result highlights a fundamental weakness of non-parametric models. While parametric models aim to provide a simplified description of the world and therefore be used for multiple purposes, non-parametric models tend to be task-oriented and can not be trivially applied to multiple goals without rethinking the model itself. 


\textit{Fourth}, throughout our analysis, we notice a steep decline in the parametric models' performance from the 2008 financial crisis to the present, which could be due to alleviated intermediary constraints during the crisis and the new financial regulation that followed \citep[see e.g.][]{chen/joslin/ni:19, haynes/etal:19}. The parameters' in-sample fit, the out-of-sample predictive power, and the hedging performance all deteriorate significantly. This observation suggests that a potential structural change might have impacted those pricing models' performances. Furthermore, we observe that the HM performance relative to that of the BDJM have decreased significantly in recent years, implying that jump risk has become a more critical pricing factor in the options market. Notice that frequent re-estimation of the models is needed to uncover these patterns. Indeed, when the model's parameters are selected to minimize the mean squared error on the whole sample, the in-sample error will mechanically be smoothed across time, while out-of-sample analysis cannot be measured.

From a methodological perspective, this paper makes two contributions to the existing literature. First, we present \textit{deep structural estimation} and discuss how to use it in practical applications. We demonstrate how to adopt the deep neural network's architecture, the neurons' activation functions, the training procedure, and the simulated training-sample to a given model's complexity---that is, the number of parameters and states, as well as the non-linearity of the model's response surface. Second, we demonstrate the usefulness of cheap-to-evaluate surrogate models in controlled environments. With simulated data, we show that \textit{deep structural} estimation can swiftly and very precisely estimate the parameters of complex structural models.

The remainder of this article is organized as follows. Section~\ref{sec:Literature} provides a very brief overview of the related literature. In Section~\ref{sec:methodology}, we present the \textit{deep structural estimation} methodology and discuss how to apply it to option pricing models. Section~\ref{sec:Result} discusses the results of our numerical experiments. In section \ref{sec:result_simulation}, we discuss in particular the performance of the surrogate in a controlled environment, whereas section \ref{sec:confront_method_to_data} confront our methodology in the context of real data. Section~\ref{sec:Conclusion} concludes.

\section{Related literature}
\label{sec:Literature}

Our paper contributes to three strands of literature: (i) methods for constructing surrogate models in general and their application to high-dimensional models in economics and finance; (ii) applications of deep learning in finance and economics; and (iii) empirical option pricing.

Model estimation, calibration, and uncertainty quantification can be daunting numerical tasks because of the need to perform sometimes hundreds of thousands of model evaluations to obtain converging estimates of the relevant parameters and converging statistics \citep[see, e.g.,][among others]{fernandez-villaverdeetal_2016,NBERw27715,10.1093/ectj/utaa019,10.1093/ectj/utaa005}. To this end, a broad strand of literature in engineering, physics \citep[see, e.g.,][]{TRIPATHY2018565}, but also in finance and economics has long tried to replace expensive model evaluations that suffer from the curse of dimensionality with cheap-to-evaluate surrogate models that mitigate the said curse.\footnote{Over the course of the past two decades, there have been significant advancements in the development of algorithms and numerical tools to compute global solutions to high-dimensional dynamic economic models (see~\cite{Maliar2014325} for a thorough review). This strand of literature is loosely related to the work presented here in that also in the context of computing global solutions to dynamic models, high-dimensional functions have to be approximated repeatedly.} \cite{HEISS200862} for instance, approximated the likelihood by numerical integration on Smolyak sparse grids, whereas~\cite{scheidegger_treccani_2018} applied adaptive sparse grids\footnote{see, e.g.,~\cite{brumm_scheidegger_2017,Pfluger2010SpatiallyProblems}, and references therein for more details on adaptive sparse grids.} to approximate high-dimensional probability density functions (PDFs) in the context of American option pricing. \cite{scheideggerbilinois_2019} propose a method to carry out uncertainty quantification in the context of discrete-time dynamic stochastic models by combining the active subspace method~\citep{Constantine_2015} with Gaussian processes to approximate the high-dimensional policies as a function of the endogenous and exogenous states as well as the parameters.
~\cite{NN_estimate_2019} introduce a framework called \textit{Calibration Neural Networks} to calibrate financial asset pricing models and provide numerical experiments based on simulated data.~\cite{2020arXiv200706169K} propose a simulation-based estimation method for economics structural models using a generative adversarial neural network. 

A paper that is relatively close to ours is \cite{norets_2012}, which extends the state-space by adding the model parameters as ``pseudo-states'' to efficiently estimate finite-horizon, dynamic discrete choice models. He uses shallow artificial neural networks to approximate the dynamic programming solution as a function of parameters and state variables prior to estimation. The main difference in our methodology is the application of deep neural networks. Compared to shallow neural networks and some of the other popular approximation methods, the benefits that deep neural nets offers include approximating power, alleviating the curse of dimensionality by reducing the amount of training data required, and the ability to make efficient use of GPUs and big data. In particular, we show that a deep network can reproduce the parameter values significantly more accurately than shallow networks, which is crucial for structural estimation. We also confront our surrogates with real data in the context of empirical option pricing.


Secondly, our paper is part of the emergent literature on applications of deep learning to economics and finance. In the seminal work of~\cite{hutchinson1994nonparametric}, shallow neural nets are used for pricing, and hedging derivative securities.~\cite{chen1999improved} established improved approximation error rate and root-mean squared convergence rate for nonparametric estimation of conditional mean, conditional quantile, conditional densities of nonlinear time series using shallow neural networks and obtained root-n asymptotic normality of smooth functionals. \cite{chen2009land} used neural networks to solve habit-based asset pricing models. More recently,~\cite{Farrell_2021} derive non-asymptotic high probability bounds for deep feed-forward neural nets in the context of semiparametric inference without a pseudo-states approach.~\cite{doi:10.1080/14697688.2019.1571683} presented a methodology called \textit{deep hedging}, which uses deep neural networks and reinforcement learning to compute optimal hedging strategies directly from simulated data with transaction costs.~\cite{didisheim_et_al_2020} introduced the concept of \textit{deep replication} to extract the implied risk aversion smile from options data.~\cite{chen/pelger/zhu:19} use deep neural networks to estimate an asset pricing model for individual stock returns that takes advantage of the vast amount of conditioning information.~\cite{beckeretal_2018} introduce \textit{deep optimal stopping} for pricing Bermudan options. In contrast,~\cite{azinovic_et_al_2019,RePEc:pen:papers:19-015,villavalaitis_2019,maliaretal_2019,duarte_2018,JFV_DL_2020} apply various formulations of deep neural networks to solve a broad range of high-dimensional dynamic stochastic models in discrete and continuous-time settings, but do not deal with estimation. We relate to this strand of literature in that we apply neural networks to tackle problems in finance. In particular, and to the best of our knowledge, we are the first to show that very deep neural nets combined with the Swish activation function~\citep{Ramachandran2017SwishAS} are necessary to construct high-dimensional surrogate models to estimate dynamic models in finance.

Thirdly, in the option pricing application, we study two popular structural models in the literature, the stochastic volatility model of \cite{heston1993closed}, and the double-exponential jump-diffusion model extended from \cite{bates1996jumps}. Our method allows us to re-estimate the structural parameters and hidden states at high frequency and compare the models' out-of-sample performances at various time horizons. We apply the test statistic developed by \cite{andersen2015parametric} to investigate the stability of the risk-neutral dynamics of the two models. 
\citet{CHRISTOFFERSEN2004291} also compare the pricing performance of a ``standard'' Practitioner's Black-Scholes (PBS) model against the Heston model. They emphasize the importance of using the same loss function to estimate and evaluate the models, and they find that in doing so, the PBS model outperforms the Heston model out-of-sample. Our finding of the superior hedging performance of structural models echoes the finding of \cite{SCHAEFER20081} in the corporate bond market, which show that structural credit models are informative about out-of-sample hedge ratios even though they might imply large pricing errors. 




\section{Methodology}
\label{sec:methodology}
In this section, we introduce our \textit{deep structural estimation} methodology. Furthermore, we briefly discuss the two option pricing models to which we apply our technology to, that is to say, the Heston~\citep{heston1993closed} and the Bates model~\citep{bates1996jumps}.

To do so, we proceed in the following steps: First, we introduce the concept of \textit{deep structural estimation} in section~\ref{sec:methodology_math}. Second, we discuss in section~\ref{sec:methodology_benefits} the general benefits of using deep neural networks in the context of surrogate models. Third, we address the specifics of our option pricing applications in~\ref{sec:methodology_option_pricing_models}. 
Fourth, we elaborate in section~\ref{sec:deep_struct_est} on the deep neural network's architecture and the training procedure applied in our applications.  
Besides, we describe in section~\ref{sec:OOS} how to assess out-of-sample performance in our numerical experiments. Finally, section~\ref{sec:rf_presentation} briefly introduces a non-parametric benchmark---random forests---to which we confront the structural results with.


\subsection{Deep structural estimation}
\label{sec:methodology_math}

Consider an economic model that can be represented by some function $f:\mathbb{R}^m\rightarrow\mathbb{R}^k$ mapping $m$ input variables into $k$ potentially observable variables. The input dimensionality $m$ can be decomposed into $\omega$ observable and time-varying states, $h$ hidden states, and a set of $\theta$ parameters, with $m=\omega+h+\theta$. 
More formally, 
\begin{equation}
f(\Omega_t, H_t | \Theta) = y_t,
\label{eq:econ_fun}
\end{equation}
where $\Omega_t$ is a vector of dimension $\omega$ containing the observables states, $H_t$ is a vector of dimension $h$ comprising the hidden states, $\Theta$ is a vector of dimension $\theta$ containing model parameters, and $y_t$ is a vector of dimension $k$ comprising the predicted quantities of interest. 

The objective now is to replace the true function $f(\cdot)$ that potentially might be expensive to evaluate by a numerically cheap-to-evaluate surrogate model $\hat{f}(\cdot)$. That is, 
\begin{equation}
\hat{f}(\Omega_t, H_t, \Theta) = \hat{f}(X_t) = y_t,
\label{eq:surrogate}
\end{equation}
where $X_t$ is represented by vector of dimension $m$ containing the observable and the unobservable states, as well as the model parameters as pseudo-states,
\begin{equation}
X_t = [\Omega_t, H_t, \Theta]^T.
\end{equation}
The core idea of \textit{deep structural estimation} is to use deep neural networks to construct the said surrogate $\hat{f}(\cdot)$ to accurately approximate the true function $f(\cdot)$, including its gradients, which helps with extreme estimators such as GMM \citep[][]{Hansen1982Econometrica}. Compared to some of the other popular approximation methods \citep[see][for a survey]{judd:96}, the benefits that DNN offers include approximating power \citep[see e.g.,][]{hornik1989multilayer}, alleviating the curse of dimensionality by reducing the amount of training data required, and the ability to make efficient use of GPUs and big data.




\subsubsection{Creating the surrogate model}
\label{sec:create_surrogate}

\begin{figure}[t!]
\centering     
\includegraphics[width=80mm]{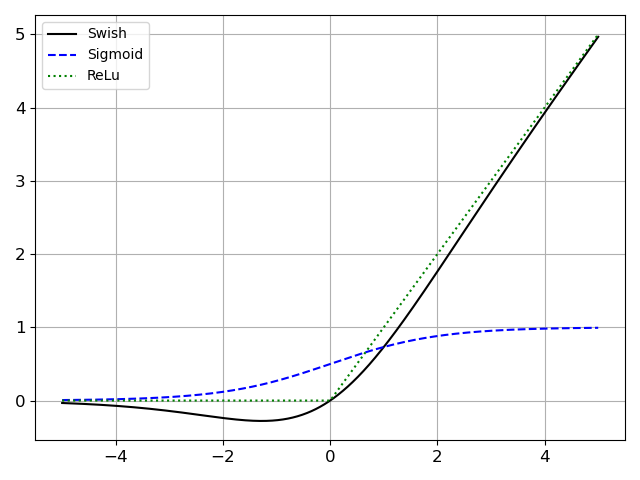}
\caption{This figure depicts the Swish, ReLu, and the Sigmoid activation functions.}
\label{fig:activation}
\end{figure}

We use deep neural networks to create the said surrogate model. Neural networks are universal function approximators \citep[see e.g.,][]{hornik1989multilayer}) that consist of stacked-up layers of neurons.
A given layer $i$ takes a vector $I_i$ of length $m$ as an input, and produces a vector $O_i$ of length $n$ as output, that is, 
\begin{equation}
O_i = \sigma\left(
W_i I_i+ b_i
\right),
\label{eq:output}
\end{equation}
where $W_i$ is a matrix of size $(m,n)$ that consists of a-priori unknown entries; $b_i$ is a vector of weights of length $n$; $\sigma(\cdot)$ is a non-linear function applied element-wise and that is commonly termed an \textit{activation function}. 
For a neural network that consists of $L$ layers, $I_1$ represents the network's inputs, whereas $I_i=O_{i-1} \forall i\in[2, L]$, and $O_L$ represents the output of the last layer. 
Popular choices for the activation function $\sigma(\cdot)$ includes rectified linear unit (Relu), $\sigma(x)=max(x,0)$, which is often preferred over the sigmoid function, $\sigma(x)=\frac{1}{1+\mathrm{e}^{-x}}$, as it does not suffer from the vanishing gradient problem~\citep{goodfellow2016deep}. However, in our numerical experiments below, we use the more recent \textit{Swish}\footnote{It has been shown empirically that the performance of very deep neural nets in combination with \textit{Swish} activation functions outperform  architectures that rely on ReLu (see, e.g.,~\cite{TRIPATHY2018565}).} activation function~\citep{Ramachandran2017SwishAS}, which is given by
\begin{equation}
\sigma(x)=\frac{x}{1+\exp (-\gamma x)},
\end{equation}
where $\gamma$ is either a constant or a trainable parameter, which can be viewed as a smooth version of the ReLu function (see figure \ref{fig:activation}). The \textit{swish} activation function possess two qualities that make it appropriate to the surrogate application: 1) unlike the \textit{sigmoid}, it does not suffer from the vanishing gradient problem and, as our application shows, complex structural models requires surrogate networks with a deep architecture, and 2) unlike the \textit{ReLu} activation function, the \textit{swish} is smooth, which means that the gradients of the trained surrogate model will also be smooth across the state-space. This latter property substantially enhances the performance when the gradients of the surrogate are need (e.g., for estimation).

Next, we discuss how to optimally determine the hyper-parameters of the deep neural network such that it can be efficiently used as a surrogate model of $f(\cdot)$ (cf. equation~\eqref{eq:econ_fun}). To do so, let $\phi(X|\theta_{NN})$ be a neural network that consists of $L$ layers. The said network takes the vector $X$ of dimension $\omega+h+\theta$ as an input, and generates an output of dimension $k$. $\Theta_{NN}$ denotes a flattened vector containing all the \textit{trainable} parameters of the neural network, that is, the components of the matrix $W_i$ and the vector $b_i$ for all layers $i=1,2,...,L$.
$\phi(X|\theta_{NN})$ is an acceptable surrogate of the model $f(\cdot)$ if 
a convergence criterion such as 
\begin{equation}\label{equ:conditions_surrogate}
\left(
\phi(X_t|\theta_{NN})- f(\Omega_t, H_t | \Theta)
\right)^2
<\varepsilon,
\end{equation}
for all economically relevant values of $\Omega_t$, $H_t$, and $\Theta$, and with $\varepsilon$ being some small positive constant representing is met. 
Thus, one somehow needs to determine the parameters $\Theta_{NN}$, which satisfy equation~\eqref{equ:conditions_surrogate}. To do so, we define a training set, that is, data used to train the neural network---composed of pairs $\tilde{X}_i, y_i$, and where $y_i = f(\tilde{X}_i)$. 


Now, we populate the by parameters enhanced state-space of dimension $m$ with sample points $\tilde{X}_i$ by drawing them from a multivariate uniform distribution. To this end, we first need to define the minimum and the maximum for every state and parameter that are economically meaningful (based on expert knowledge or based on mathematical conditions). Thus, for every state $x^{(j)}_t$ that populates the vector $X_t$, let $\underline{x}^{(j)}_t$ be the minimum value acceptable, and let $\bar{x}^{(j)}_t$ be the maximum admittable value,
\begin{align} 
\underline{X} &= [\underline{x}^{(1)},\underline{x}^{(2)},...,\underline{x}^{(\omega+h+\theta)}], ~~
\bar{X} = [\bar{x}^{(1)},\bar{x}^{(2)},...,\bar{x}^{(\omega+h+\theta)}].
\end{align}
Next, we can define
\begin{equation}\label{equ:random_point_gen}
\tilde{X}_i = \underline{X} + R (\bar{X}-\underline{X}),
\end{equation}
where $R=[r^{(1)},r^{(2)},...,r^{(\omega+h+\theta)}]$\citep{1456693}.\footnote{The vector $r$ can be drawn from any distribution appropriate to the particular context. In this paper, we use a simple uniform distribution for all states and parameters.} Consequently, one can now generate a training sample of size $N$ by drawing $N$ random vectors $\tilde{X}_i$, and querying the original model (cf. equation~\eqref{eq:econ_fun}) $N$ times to obtain $y_i=f(\tilde{X}_i)$.\footnote{Note that with complex structural models and large $N$, this operation can be computationally costly. However, this expense poses a unique fixed cost in the sense that the researcher will need to spend these computational costs only once. After that, in practice, he can benefit from orders of magnitude of speedup when querying the surrogate model rather than evaluating the original function (cf. section~\ref{sec:Result}). Furthermore, the generation of training data is embarrassingly parallelizable and thus consumes only little human time. As an illustration, consider an estimation of the BDJM's parameters using the original model function. This optimization took roughly 40 minutes. On the same hardware, we reduced the optimization time to less than 1 second with the \textit{deep surrogate estimation} technology. Please keep in mind that these numbers are only here to illustrate the potential time gain. Indeed, both times could probably be reduced significantly with expert knowledge and appropriate code optimization. All tests were conducted on the ``Piz Daint," a Cray XC$50$ system that is installed at the Swiss National Supercomputing Centre. Its compute nodes are equipped with a $12$-core Intel(R) Xeon(R) E$5$-$2670$ with $64$GB of DDR$3$ memory.}
Given this set of training data, we can now train the neural network $\phi(X|\theta_{NN})$ by minimizing the mean absolute error on the training sample, that is,
\begin{equation}\label{equ:surrogate_minimization}
\theta_{NN}^{*}=\underset{\theta_{NN}}{\arg \min } 
\frac{1}{N} \sum_{i=1}^{N} 
|\phi(\tilde{X}_i|\theta_{NN})-y_i|.
\end{equation}
To perform this minimization, we need to access the neural network's gradients. This can be done very efficiently thanks to the backpropagation algorithm, which we can view as a recursive application of the standard chain rule~\citep{chauvin1995backpropagation}. Note that, unlike other numerical differentiation schemes, backpropagation is exact. We compute the gradient to update the network's parameters on mini-batches in a process known as stochastic gradient descent. We can either use stochastic gradient descent until convergence or stop earlier to avoid overfitting. However, in the case of \textit{deep structural estimation} presented here, early stopping is unnecessary thanks to the fact that our target $y_i$ can safely be assumed to be noise-free and that $N$ can be made arbitrarily large. 

To test whether or not the neural network is surrogate model of acceptable quality, we can generate an additional sample of pairs $\tilde{X}_j, y_j$, for $j=1,...,J$: the validation sample with which we compute the validation error, i.e.,
\begin{equation}\label{equ:validation_error}
\frac{1}{N} \sum_{j=1}^{J} 
|(\phi(\tilde{X}_i|\theta_{NN})-y_i)|.
\end{equation}
If the validation error is smaller than some prescribed $\varepsilon$, we define the neural network as an acceptable surrogate of the model $f(\cdot)$. 


\subsubsection{Using the surrogate model for structural estimation}
\label{sec:using_surrogate}

Let us assume that we observe a time series of economic data from time $t=1$ to $t=T$. At each time step, we observe $N_t$ target states and their corresponding observable states,
\begin{equation}
\begin{array}{rcl} 
\hat{Y}_t &=& [\hat{y}_t^{1},\hat{y}_t^{2},...,\hat{y}_t^{N_t}] ,\\
\hat{\Omega}_t &=& [\hat{\Omega}_t^{1},\hat{\Omega}_t^{2},...,\hat{\Omega}_t^{N_t}],
\end{array}
\end{equation}
where $\hat{Y}_t$ is a matrix of size $(k, N_t)$ that consists of all the observed states $\hat{y}_t^{j}$ at time $t$ which the true model $f(\cdot)$ attempts to explain. Furthermore, $\hat{\Omega}_t$ represents a matrix of size $(\omega,N_t)$ that is composed of the corresponding observable state of the economy. We wish now to use the said data to estimate the model's parameters, $\Theta$, as well as the hidden state of the economy: 
\begin{equation}
\hat{H}_t = [\hat{H}_t^{1},\hat{H}_t^{2},...,\hat{H}_t^{N_t}],
\end{equation}
where $\hat{H}_t$ is a matrix of size $(h,N_t)$, for all $t=1,...,T$. With the \textit{deep structural estimation} technology, we can now directly and swiftly solve the following minimization problem:  
\begin{equation}\label{equ:surrogate_minimization}
\hat{H}_1^*, \hat{H}_2^*,...\hat{H}_T^*, \Theta^* =\underset{\hat{H}_1, \hat{H}_2,...\hat{H}_T, \Theta}{\arg \min } 
\frac{1}{T}
\frac{1}{N_t} \sum_{\tau=1}^{T} \sum_{i=1}^{N_t} 
(\phi(\left[\hat{\Omega}_{\tau}^{i}, \hat{H}_{\tau}^{i}, \Theta \right]|\theta_{NN}^*)-\hat{y}_\tau^{(i)})^2.
\end{equation}
In our numerical experiments below (cf. section~\ref{sec:Result}), we apply the BFGS optimization algorithm~\citep[see, e.g.,][]{nocedal2006numerical}. The BFGS procedure relies on a correct estimation of the function's gradient, which we can obtain cheaply and precisely thanks to the backpropagation algorithm.\footnote{Note that contemporary APIs such as \textit{TensorFlow 2.x} provide an implementation of machine learning algorithms, including deep neural networks and classical optimization algorithm like the BFGS in a single framework. Therefore, \textit{deep structural estimation} was implemented within this API to smoothly leverage  the availability of gradients with a well-tuned BFGS algorithm and the available parallelization to harvest the computing power of contemporary GPUs.} 
Note that even without the \textit{deep structural estimation }, we could reach the desired results by solving, 
\begin{equation}\label{equ:brute_force_minimization}
\hat{H}_1^*, \hat{H}_2^*,...\hat{H}_T^*, \Theta^* =\underset{\hat{H}_1, \hat{H}_2,...\hat{H}_T, \Theta}{\arg \min } 
\frac{1}{T}
\frac{1}{N_t} \sum_{\tau=1}^{T} \sum_{i=1}^{N_t} 
(f(\hat{\Omega}_{\tau}^{i}, \hat{H}_{\tau}^{i}| \Theta)-\hat{y}_\tau^{(i)})^2.
\end{equation}
However, unlike equation~\eqref{equ:surrogate_minimization}, we would have to compute the model's gradient directly, which would be significantly more costly for three reasons. First, unless the model's gradient can be derived analytically, one would have to estimate the gradients through a numerical differentiation scheme, requiring multiple costly evaluations of the model. Second, even a single estimation of the model can be computationally very expensive for complex structural models. 
Third, depending on the structural model, it may be impossible to accelerate the gradient's computation. Furthermore, even if we can parallelize the calculation of the model's gradient on modern GPUs, the actual implementation of this parallelization will be a daunting task requiring time and expert knowledge.


\subsection{The benefits of deep structural estimation}
\label{sec:methodology_benefits}


When dealing with contemporary structural models in finance or economics, the computational costs can be a limiting factor. The method we described in the previous section provides a trade-off for the researcher. One pays a one-time, potentially relatively high, upfront aggregate computational cost\footnote{Note that parallelizing the generation of the training set is trivial. Thus, although the overall computational cost can be "large" in node hours, it can be distributed across the compute nodes of contemporary high-performance computing hardware, thus reducing the runtime by orders of magnitude.} to reduce the marginal cost of additional model estimations by orders of magnitude. This trade-off provides two main advantages. First, the cheap marginal cost of calibration allows a thorough out-of-sample and parameter stability analysis. Second, the surrogate model can easily be stored and shared with other users. With \textit{deep structural estimation}, we can create a library of models that can significantly improve research quality through easier access to prior work for model comparisons, meta-analysis, and recalibration on new data.

\subsection{The option pricing models}
\label{sec:methodology_option_pricing_models}


We now briefly summarize the asset pricing models for which we build surrogate models, namely the stochastic volatility model (HM) and the double-exponential jump-diffusion model (BDJM). Both models specify a stochastic process for the underlying asset $S$ under a risk-neutral probability measure $\QQ$, which is guaranteed to exist by no-arbitrage. The prices of the European options are equal to the risk-neutral expected present values of the terminal payoffs discounted at the risk-free rate. 

While there are more sophisticated option pricing models in the literature \citep[for example,][]{duffie/pan/singleton:00, bates:00,pan:02, andersen2015parametric}, we choose the HM and the BDJM model as examples mainly for their transparency: both models are single-factor, with the latter adding a jump component relative to the former. A comparison between the two allows us to focus on the importance of jumps for option pricing.

\subsubsection{The Heston model}
\label{sec:heston_model}

In the Heston model, under measure $\QQ$, the stock price $S_t$ follows the process:
\begin{subequations}
\begin{align}
\frac{dS_{t}}{S_{t}} &= (r-d) dt + \sqrt{v_{t}} dW_{1,t}, \\ 
dv_{t} &= \kappa (\theta - v_{t}) dt + \sigma \sqrt{v_{t}} dW_{2,t}. 
\end{align}
\end{subequations}
Here, $r$ is the (constant) instantaneous risk-free rate and $d$ is the dividend yield. $v_{t}$ is the (diffusive) spot volatility, which follows a Feller process under $\QQ$ with speed of mean reversion $\kappa$, long-run mean $\theta$, and volatility parameter $\sigma$. $W_{1,t}$ and $W_{2,t}$ are two standard Brownian motions under $\QQ$ with $\text{corr}(W_{1,t}, W_{2,t}) = \rho$. 

For an European option with maturity $T$ and strike price $K$, the HM model involves six parameters, that is, $\Theta_{HM}=[r, d, \kappa,\theta,\sigma,\rho]^T$, one hidden state, $H_{HM}=[v_t]$, and three observable state variables, $\Omega_{HM}=[S_t, K, T-t]^T$. 
In the estimation, we will treat two of the parameters, interest rate $r$ and dividend yield $d$, as observables, and estimate the remaining four parameters along with the hidden state $v_{t}$ from the options panel.



\subsubsection{The Bates model with double exponential jump-diffusion process}


The Bates model \citep{bates1996jumps} extends the Heston model by adding jumps in the stock price. Under $\QQ$, the stock price follows
\begin{subequations}
\begin{align}
\frac{dS_{t}}{S_{t^{-}}} &= (r-d-\lambda m) dt +\sqrt{v_{t}} dW_1 + dZ_t, \\ 
dv_{t} &= \kappa (\theta - v_{t}) dt + \sigma \sqrt{v} dW_2,
\end{align}
\end{subequations}
%
where the key difference from the HM model is $Z$, which is a pure jump process with arrival intensity $\lambda$ and the log jump size has distribution $\omega$. Different from \cite{bates1996jumps}, who assumes $\omega$ is normal, we model the log jump size with an asymmetric double exponential distribution, 
\begin{align}
\omega(J)&= p\frac{1}{\nu_u}e^{-\frac{1}{\nu_u}J} 1_{\{J>0\}} + (1-p) \frac{1}{\nu_d}e^{\frac{1}{\nu_d}J} 1_{\{J<0\}}, ~~ \text{with} ~~ p \geq 0,
\end{align}
where $\nu_u, \nu_d > 0$ are the average log jump size for positive and negative jumps, respectively. Under this specification, $m = \frac{p}{1-\nu_u} + \frac{1-p}{1+\nu_d} -1$. The double-exponential assumption allows for heavier tails in jumps than the normal distribution, a feature supported by the non-parametric option-based evidence in \cite{Bollerslev/Todorov:2011}.

In summary, BDJM consists of 10 parameters (of which 8 will be estimated, excluding $r$ and $d$), $\Theta_{BDJM}=[r, d, \kappa, \theta, \sigma, \rho, \lambda, \nu_u, \nu_d, p]^T$, plus the same hidden state $H_{BDJM}=[v_t]$ and observable states $\Omega_{BDJM}=[S_t, K, T-t]^T$ as the HM.\footnote{For both structural models, we apply the state-of-the-art option pricing library QuantLib (\url{https://www.quantlib.org}) to compute prices and the corresponding BSIV.}
 


\subsection{Deep structural estimation for option pricing}
\label{sec:deep_struct_est}


We now detail the specific training sample, network architecture, and the training procedure we use to create surrogate models for the 10-dimensional HM and 14-dimensional BDJM. 

\subsubsection{Training sample}

Before starting with any numerical computation, we exploit basic properties in the two models to reduce the effective dimensionality. To  decrease the original dimensionality by one, we combine the strike price $K$ and spot price $S$ into a measure of moneyness, i.e.,
\begin{equation}\label{equ:moneyness_definition}
\hat{K} = 100 \frac{K}{S}.
\end{equation}
Next, we further diminish the dimensionality of the original models by
collapsing the option type dimension through the call-put parity. Instead of providing the option types (call or put) as an input of the surrogate, we use the call-put parity to transform all the put prices in the sample into call prices.\footnote{Note that the choice of transforming puts into calls instead of puts into calls was arbitrary. We can obtain the same performance with a put-only surrogate model.}


Furthermore, we replace the price of the options with the corresponding BSIV. Thus, instead of having surrogate models that produce prices that we can then transform into BSIV, the surrogates directly predict the BSIV.

We populate the training sample $\tilde{X}_i, y_i$ as defined in section~\ref{sec:methodology_math}. To do so, we first set the minimum and maximum values for each state variable and all parameters. The tables~\ref{table:hm_range} and~\ref{table:bdjm_range} in appendix \ref{sec:appendix_simulated} list those ranges both for the HM and BDJM, respectively. We chose those values based on a mixture of mathematic rules, expert knowledge, but also based on trial and error. For example, the intensity of the Poisson process $\lambda$ is, by definition, positive. We know that the correlation between volatility shocks and price shocks $\rho$ is only negative in practice. Finally, we observe that a large value of $\kappa$ is often necessary to obtain the best in-sample fit. Therefore, we use a relatively large range of possible $\kappa$ value to accommodate this fact. 

The only state variable in $\tilde{X}_i$ which we do not draw from a uniform distribution is $\hat{K}$. Indeed, the standardized strike $\hat{K}$ is an imperfect substitute for moneyness, as moneyness is best represented as a function of strike price and time-to-maturity. Following~\cite{andersen2017short}, we define a moneyness indicator $m$ as 

\begin{equation}\label{equ:m_coeff}
m=
\frac{\ln \left(\hat{K} / F_{T}\right)}
{\sqrt{T} \sigma_{a t m}},
\end{equation}
where $F_{T}$ is the forward for transaction up to the option's maturity, and $\sigma_{atm}$ is the average BSIV of at-the-money options throughout the sample (0.19). Instead of drawing $\hat{K}$, we draw a random moneyness value $m$, and invert equation \eqref{equ:m_coeff} to get the corresponding $\hat{K}$.

Finally, we need to define the size of the training sample $N$, that is, the number of points populating the pseudo-state-space we use to calibrate the neural network. We choose $N=10^{8}$ for the HM, and $N=10^{9}$ for BDJM.\footnote{Note that while those numbers read like big numbers, they actually only sparsely populate the state space. Consider a naive discretization scheme, where one places $N$ points along one axis. The naive generalization $d$ dimensions would yield $N^d$ points. Consequently, those $N=10^{8}$ points in the case of the BDJM would translate to $N\approx4$, which could be considered a low resolution of the function to be approximated with traditional, Cartesian grid-based methods (see, e.g.,~\cite{Press:2007:NRE:1403886}).} We chose these numbers through trial and error by training multiple surrogate models on increasingly larger training samples until the result produced in our simulation reached a satisfying performance level.

\subsubsection{The deep neural network's architecture}
\label{sec:methodology_net_architecture}
We applied a network of 6 hidden layers to generate the HM surrogate, each of which was composed of 400 neurons with a \textit{Swish} activation function. This architecture yields a total of 806,001 trainable parameters. With the BDJM surrogate, we use a deep neural network with seven hidden layers, 400 neurons each, and \textit{Swish} activation functions. The latter architecture corresponds to a total of 967,201 trainable parameters. Our trial and error approach to determine the optimal architecture---as common when working with deep neural networks---suggests that all things being equal, the network has to be deeper with additional complexities (such as the addition of states and model parameters) to keep the surrogate performance at a constant level.\footnote{Note that we also tried to increase the network's width, that is, to add more neurons in each layer. This procedure led to a considerably increased training time but did not significantly increase the surrogate performance.}

\subsubsection{The deep neural network's training procedure}

We run a mini-batch stochastic gradient descent algorithm with batches of size $256$ to determine the neural network's parameters. In particular, we applied the ADAM algorithm with an initial learning rate of $0.5*10^{-4}$. As optimization criteria, we minimize the mean absolute error (cf. equation~\eqref{equ:surrogate_minimization}). 

We run the optimization algorithm for $15$ epochs, that is, we use mini-batches of size $256$ until we have used the whole data-set 15 times. After each epoch, we save the model and use a validation set of 10,000 points to estimate the surrogate model's performance. At the end of the procedure, we use the network's parameters after the epoch with the lowest validation error. 


\subsection{Out-of-sample pricing errors}
\label{sec:OOS}


To measure out-of-sample pricing errors, we estimate the structural option pricing models' parameters and hidden states at some given time $t$, and use these states and parameters to predict options' BSIVs at time $t+\tau$, where $\tau$ is the forecasting horizon defined in numbers of business days. We make the predictions assuming we can see the observable states of time $t+\tau$, that is, the options' maturity, the options' moneyness, and the risk-free rate. Formally, we define the out-of-sample prediction for each option $i$ as 

\begin{equation}
\hat{y}_{i}= \phi( 
\left[\hat{\Omega}_{\tau}^{i}, \hat{H}_{0}^{i}, \Theta_{0}\right]
| \Theta_{NN}^*).
\end{equation}

Note that we do not update the state parameter $v_t$. In practice, we measure the volatility smile on the day $t$ and use it to predict the volatility smile at time $t+\tau$. As such, the performance out-of-sample can be viewed as a measure of parameter stability.


\subsection{Non-parametric benchmark}
\label{sec:rf_presentation}


To contrast the performance of the two structural option pricing models, we apply a non-parametric benchmark in the form of a random forest regressor.\footnote{A random forest works by constructing a multitude of decision trees, with the algorithm's prediction defined as mean output of each individual tree. This ensemble approach drastically reduces the risk of over-fitting. For a general introduction to random forests, see~\cite{liaw2002classification}, and for random forests applied to finance see~\cite{gu2018empirical}.}  
We choose the random forest over other non-parametric benchmarks for two main reasons: i) random forests regressors requires little to no tuning to perform well out-of-sample, ii) random forest regressors have a relatively short training time, which significantly facilitates our analysis. 
 
To train a random forest, we need to define the model's input, that is, the vector of observable states $X_i^{(R)}$ the random forest uses to predict the BSIV of option $i$---and the loss function which the random forest tries to minimize, that is,\footnote{Concerning the random forest hyperparameters, we applied the default parameters of the \textit{sklearn} library (cf.~\url{https://scikit-learn.org/stable/modules/generated/sklearn.ensemble.RandomForestRegressor.html}).}
\begin{equation}
X_i^{(R)}=[ 1, \mathbbm{1}_{C,i}, \hat{K}_i, T, \hat{K}_i T,\mathbbm{1}_{\hat{K}>100}],
\end{equation}
where $1$ is a constant term, $\mathbbm{1}_{C,i}$ is a binary indicator equal to $1$ if the contract is a call option, $\hat{K}_i$ is the moneyness measure defined in equation \eqref{equ:moneyness_definition}, $T_i$ is the option's days to maturity, and $\mathbbm{1}_{\hat{K}>100}$ is a binary indicator equal to $1$ if the option's strike is above the underling's asset price. The last two dimensions of the vector $X_i^{(rf)}$ provide redundant information, which we include to speed up the algorithm training. 

Let $R(\cdot)$ be a random forest predictor. To create a non-parametric benchmark (cf. section~\ref{sec:OOS}), we train a random forest to minimize the following loss function:
\begin{equation}
\mathcal{L}  = \frac{1}{N} \sum_{i=1}^{N} \left(
R(X_i^{(R)}) - \hat{y}_i)
\right)^2,
\end{equation}
where $N$ is the number of options in the training sample, and $\hat{y}_i$ is the option's BSIV. 


Note that the random forest minimizes the mean squared error, just as the surrogate models discussed above. In addition, our non-parametric benchmark does not use any additional information other than the one we use to estimate the option pricing model's hidden states and parameters.

%

\section{Result}
\label{sec:Result}

This section presents the application of the \textit{deep structural estimation} technology to the HM and BDJM models. For reasons of transparency, we start in section~\ref{sec:result_simulation} by using simulated data to demonstrate the capabilities of our framework in a controlled environment. In section~\ref{sec:confront_method_to_data}, we turn our attention towards a panel of S\&P500 options data to calibrate both the HM and BDJM model at a daily frequency to perform an analysis that would be computationally (almost) infeasible without the surrogate technology. In particular, we consider and compare several definitions of out-of-sample performance to discuss the performance of the models. In each case, we compare the performance of the two options pricing models against each other and a non-parametric benchmark: the random forests. In addition, we discuss the models' parameter stability across time. Finally, we compare the models' capacity to predict the delta of individual options, that is, the options' price sensitivity to changes in the underlying asset's price.  

\subsection{Controlled test cases}
\label{sec:result_simulation}

In this section, we demonstrate the capability and versatility of the surrogate technology in a controlled environment. 
We start by first investigating the sensitivity of surrogates to their respective states and pseudo-states.
To do so, we apply equation~\eqref{equ:random_point_gen} to randomly draw a sample point. Then, we vary each state and pseudo-state in turn while keeping all other variables fixed. We compute the BSIV by querying the original option pricing model and querying the surrogate for each state vector to produce a sensitivity plot. 

Figure~\ref{fig:sensitivity_in_paper} displays the resulting sensitivities for some states and parameters from the BDJM's surrogate. The y-axis shows the BSIV as a function of a given pseudo-state from the range given in table~\ref{table:bdjm_range}. We can see that the surrogate model almost perfectly replicates the sensitivity of the target model. Furthermore, these plots highlight the high non-linearity of the BDJM's volatility surface. Taken together, these results show that neural network surrogates can reach a high degree of precision for very complex structural models. Figures~\ref{fig:sensitivity_hm},~\ref{fig:sensitivity_bate_1}, and~\ref{fig:sensitivity_bate_2} in appendix~\ref{sec:appendix_simulated} show similar sensitivity graphs for all states and parameters of both surrogates. 

\newcommand{\ww}{0.3\linewidth}
\begin{figure}[t]
\centering     
\subfigure[]{\label{fig:hest_call_pred_err}\includegraphics[width=\ww]{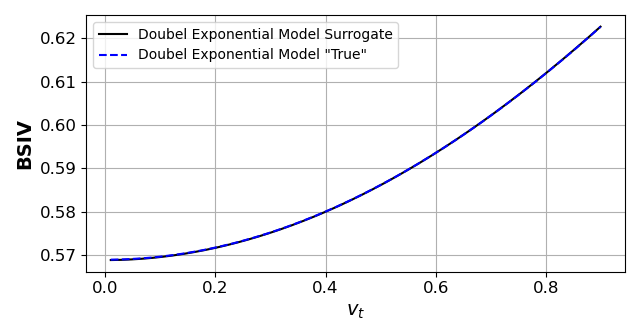}}
\subfigure[]{\label{fig:hest_call_pred_err}\includegraphics[width=\ww]{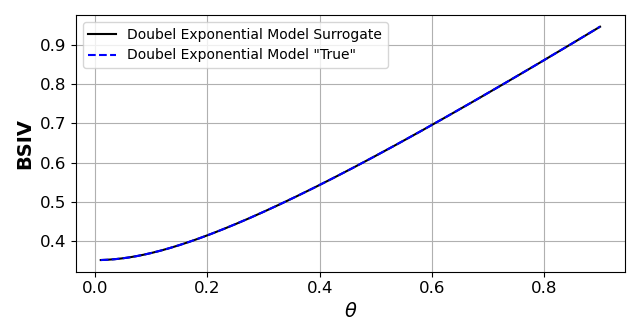}}
\subfigure[]{\label{fig:hest_call_pred_err}\includegraphics[width=\ww]{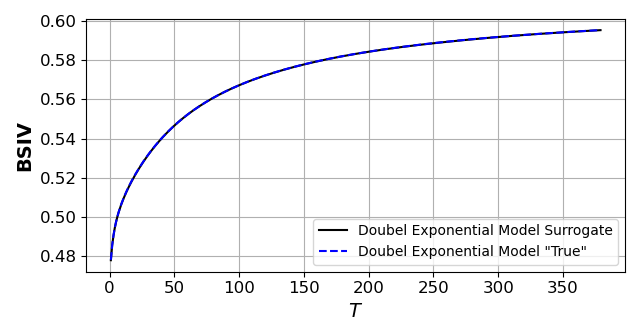}}
\subfigure[]{\label{fig:hest_call_pred_err}\includegraphics[width=\ww]{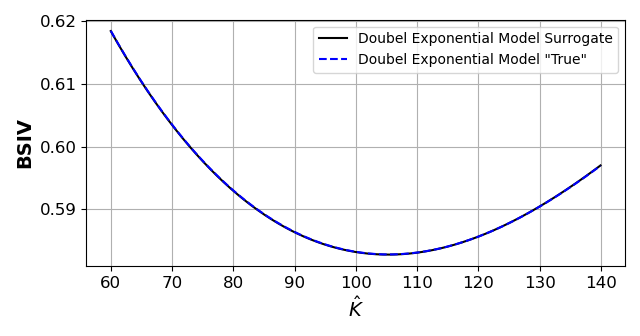}}
\subfigure[]{\label{fig:hest_call_pred_err}\includegraphics[width=\ww]{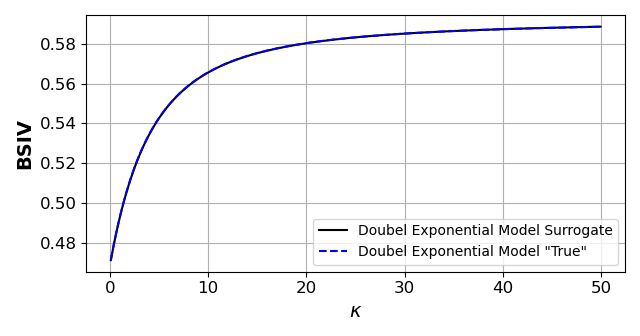}}
\subfigure[]{\label{fig:hest_call_pred_err}\includegraphics[width=\ww]{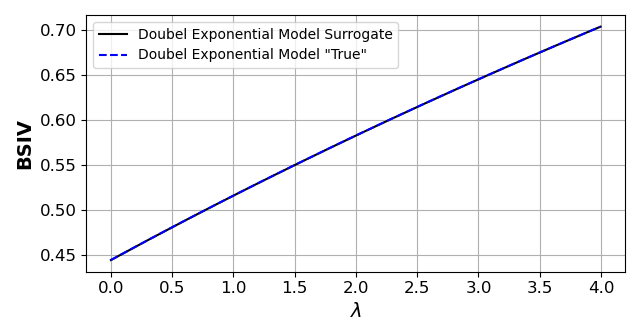}}
\subfigure[]{\label{fig:hest_call_pred_err}\includegraphics[width=\ww]{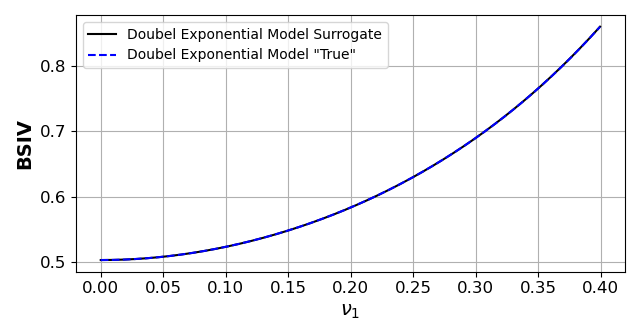}}
\subfigure[]{\label{fig:hest_call_pred_err}\includegraphics[width=\ww]{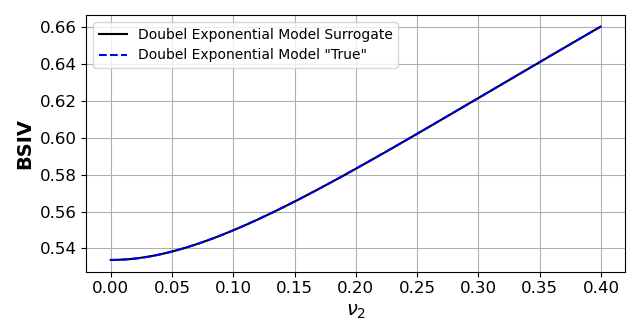}}
\subfigure[]{\label{fig:hest_call_pred_err}\includegraphics[width=\ww]{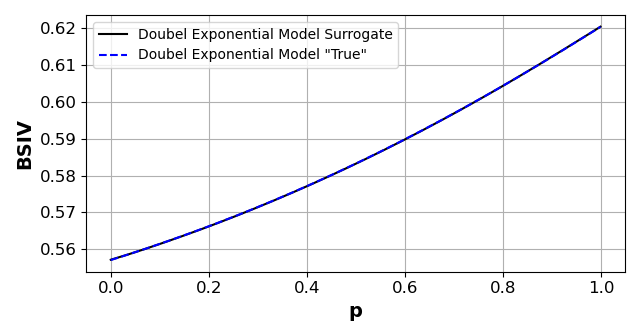}}
\caption{The figures above show the complexity of the surrogate models' volatility surface as well as the quality of the surrogate's interpolations. To do so, the figures compare the BSIVs of the surrogate against the predictions of the ``true" BDJM. We populate the state space with points, where we keep all parameters and states fixed at the mid-range of possible values: $\hat{K}=100.0,rf=0.0375, T=190, \kappa=25.05, \theta=0.455, v_t=0.455, \sigma=2.55$. On each panel, we show the BSIV predicted by the BDJM and its surrogate while varying one of the parameters or states in its admissible range (cf. table~\ref{table:bdjm_range}).}
\label{fig:sensitivity_in_paper}
\end{figure}

%

Next, to demonstrate the value of the introduced surrogate technology in the context of structural estimation, we perform the last steps described in section \ref{sec:methodology_math}. We estimate the states and pseudo-states from the data. We start out by randomly generate a cross-section of the options' BSIV by drawing one state vector $R^{true}$ (cf. equation~\eqref{equ:random_point_gen}), that is,
\begin{equation}ion
R^{true}=\left[\Omega^{true}, H^{true}, \Theta^{true}\right]^T.
\end{equation}
After that, while keeping all other parameters and states equal to their value in $R^{true}$, we populate the simulated cross-section of options by varying the moneyness and the maturity parameter, $\hat{K}$ and $T$. Thus, we create a set of state vectors $\tilde{X}_i$ with parameters and state equal to the one in $R^{true}$ except for the moneyness and maturity parameters, $T, \hat{K}$. Now, we query the option pricing model to estimate a price for the option, and we convert this price into a BSIV, $\tilde{y}_i = f(\tilde{X}_i$). 

We apply the described procedure to generate a cross-section of $N=1,000$ points and use the optimization algorithm described in section~\ref{sec:methodology_math} to solve the following minimization problem: 

\begin{equation}
\hat{H}_{1}^*, \Theta^*=
\underset{\hat{H}_{1}, \Theta}{\arg \min } 
\frac{1}{N} \sum_{i=1}^{N}\left(\phi\left(\left[\hat{\Omega}_{1}^{i}, \hat{H}_{1}^{i}, \Theta\right] \mid \theta_{N N}^{*}\right)-\hat{y}_{\tau}^{(i)}\right)^{2}.
\end{equation}

Subsequently, we define a performance measure for the model calibration. To this end, let $X^*=[\hat{H}_{1}^*, \Theta^*]^T$ be the estimated state vector of dimension $h+\theta$ containing the unobservable states, and the models' parameters are obtained through \textit{deep structural estimation}. Moreover, let $x^*_i$ be the individual states and parameters populating the vector $X^*$, that is, $X^*=[x_1^*, x^*_2, x^{*}_{\omega+h+\theta}]^T$. For each $x^*_i$, we compute the estimation error as
\begin{equation}\label{equ:pred_error_def}
e_i=
\frac{|x^{true}_{i} - x^{*}_{i}|}
{|\bar{x}_i - \underline{x_i}|},
\end{equation}
where $\bar{x}_i$ and $\underline{x_i}$ represent the maximum and minimum values in the training sample of the respective surrogate model (cf. tables~\ref{table:hm_range} and \ref{table:bdjm_range}). This error measure captures the absolute difference between the estimated state and the true value, standardized by the possible parameter range to allow for comparison across states. 

\begin{figure}[t]
\centering     
\subfigure[HM]{\label{fig:hest_call_pred_err}\includegraphics[width=80mm]{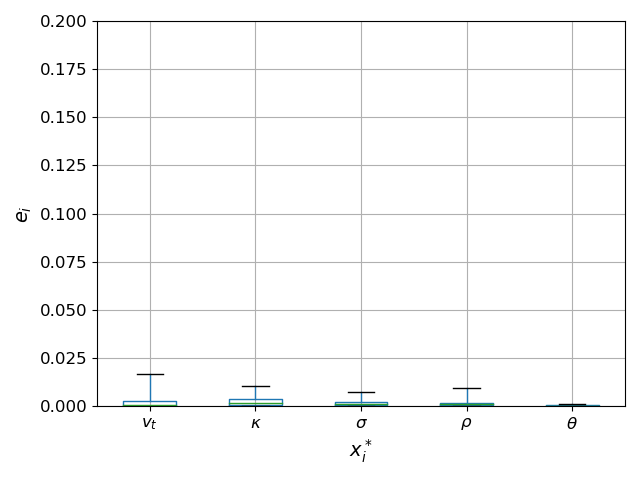}}
\subfigure[BDJM]{\label{fig:hest_put_pred_err}\includegraphics[width=80mm]{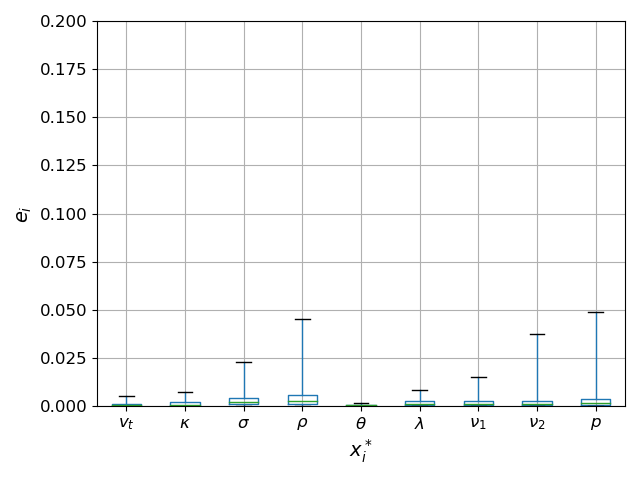}}
\caption{The figures above show that the surrogate can be used to estimate the parameters of the models on a small data sample. We simulated market days with the original pricing models and used our surrogates to estimate the parameters and states simultaneously. Above, we show the resulting prediction errors (see equation \eqref{equ:pred_error_def}) for each hidden state and parameter of both models. The green vertical line represents the median prediction error computed across 1000 simulations. The box shows the interquartile range, while the whiskers show the 1st and 99th percentile errors across the simulations.}
\label{fig:error_plots}
\end{figure}

For each surrogate model, we simulate 1,000 such cross-sections, and for each simulation, we estimate the parameters and compute the estimation error. Figure~\ref{fig:error_plots} displays the results for both models. We show for each non-observable state and parameter the standardized estimation error $e_i$ in the form of a box plot. The (green) horizontal line represents the median error. The whisker indicates the 1st and 99 quantiles, whereas the box shows the inter-quartile range across all simulations. 

Besides, we compute for every simulation the in and out-of-sample performance. The in-sample performance is defined as the MSE computed on the original cross-section $\tilde{X}_i, \tilde{y}_i$ for all $i=1,...,N$. We estimate the out-of-sample performance as the MSE on an additional set of points $\tilde{X}_j, \tilde{y}_j$ for all $j=1,...,N'$.

In both in- and out-of-sample, the mean squared pricing error is virtually $0$, with values ranging ranging from the smallest at $0.6*10^{-8}$ for the HM with call options to the largest at $0.3*10^{-5}$ for the BDJM with put options.

The results above suggest that \textit{deep structural estimation} can be used to efficiently and precisely estimate the states and pseudo-state from a simulated cross-section of options we generate with the true model's pricing function. Remember that, unlike the original models, querying the surrogate or estimating the surrogate's gradients is at a negligible computational cost. 

Before discussing the option pricing models' performances on real market data, we demonstrate the importance of the neural networks' architecture presented in section \ref{sec:methodology_net_architecture} in the surrogate models' performance. In particular, we highlight the importance of the network depth combined with the \textit{Swish} activation function. To do so, we first investigate the performance of shallow networks as surrogate models. In particular, we define a neural network with a 400 neurons layer and \textit{Swish} activation function and train it to be a surrogate for the HM call options. We apply the procedure described at the start of this section to estimate the estimation error and how well the shadow network sensitivities replicates the true model sensitivities. 
We measure the sensitivities of a surrogate model generated by a shallow neural network, as previously was done (cf. figure~\ref{fig:sensitivity_hm}). Figure \ref{fig:sensitivity_shallow} in appendix \ref{sec:appendix_simulated} displays this sensitivity comparison for the HM call surrogate. We can see that the surrogate here cannot replicate the true model efficiently. 
 
Next, we use our shallow surrogate model to compute estimation errors and produce a box-plot of those errors. Figure~\ref{fig:error_plot_shallow} in appendix \ref{sec:appendix_simulated}  displays the main findings: while the surrogate model estimates some states correctly, notably the parameter $\theta$ and the hidden state $v_t$, the error on other parameters explodes, most notably on the $\kappa$ parameters, where the median estimation error across simulations is above 13\%. 
We perform a similar analysis with a deep (6 hidden layers consisting of 400 neurons each) network with ReLU activation functions and obtained large replication errors.

These additional numerical results complete the tests for our method within a controlled environment and highlight the importance of choosing an appropriate network architecture to create efficient surrogate models. Note that, even though these \textit{bad} surrogates can not estimate the models' parameters reasonably, the prediction error of these surrogates on a validation set was still virtually 0, that is, the surrogate BSIV versus the models BSIV for a given set of parameters. With complex structural financial models, some parameters have a relatively small impact on the target state. For example, on a cross-section of options on a given day, the $\kappa$ parameter does not have a significant impact on the options BSIV. Therefore, we can reach a low prediction error while almost ignoring the effect of the $\kappa$. An appropriate network architecture can ensure that an adequate degree of precision is reached, even for marginally important parameters. Furthermore, this interesting result demonstrates that the prediction error alone is not is not a sufficiently accurate measure to assess the quality of a surrogate. 

\subsection{Confronting \textit{deep structural estimation} to market data}
\label{sec:confront_method_to_data}


\subsubsection{Data}

For the numerical following numerical experiments, we use the daily options' quotes from the \textit{OptionMetric} database. We use European call and put options on the S\&P500 index in the years between 2001 and 2019. We remove options with maturity above 250 days. Furthermore, we remove options with extreme moneyness values from the sample by keeping only options with a BS delta between 0.1 and 0.9 ($0.1\leq \Delta \leq 0.9$). The option metrics database provides, for each option, the prices, but also the BSIV as well as the BS delta at a daily frequency.  

\begin{figure}[t]
\centering     
\subfigure[Maturities in sample]{\label{fig:daily_sample_mat}\includegraphics[width=80mm]{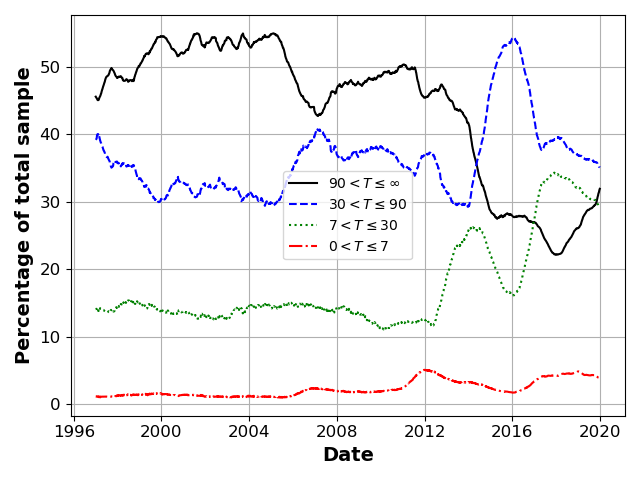}}
\subfigure[Number of options]{\label{fig:daily_sample}\includegraphics[width=80mm]{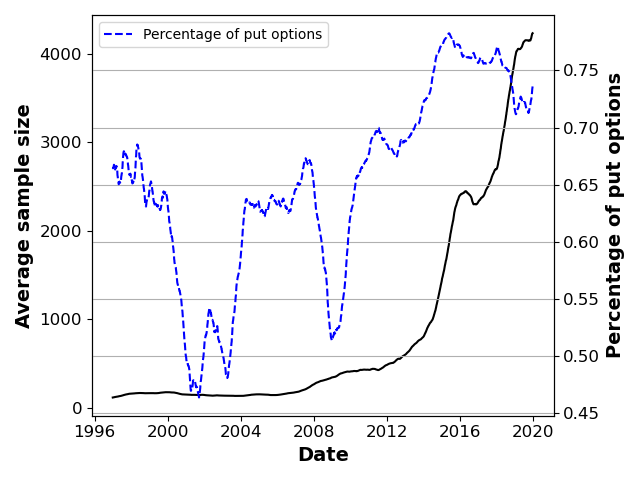}}
\caption{The two figures above show the evolution of the sample composition across time. Panel (a) shows the percentage of options in the sample across time per maturity brackets. Panel (b) shows the total number of options per day (left axis) and the percentage of put options in the sample (right axis). 
Through all plots, we smooth the results with a rolling average over the last 252 business days. 
}
\end{figure}

Figures~\ref{fig:daily_sample} summarizes our final sample size and composition. Panel (a) shows the evolution across time of the sample's composition per maturity brackets. Panel (b) shows the evolution of the sample size (left axis) and the percentage of put options in the sample (right axis). We smooth all numbers with a rolling average over the last 252 days. 


In addition, we obtain the volatility index VIX, a non-parametric estimate of jump risk in the S\&P500 index from Todorov and Andersen's website.\footnote{https://tailindex.com/volatilitymethodology.html} 
The variable $JUMP_t$ represents the weekly probability of a negative jump of at least 10\% of the S\&P500 index. 
Finally, we obtain the risk-free rate from the Fama and French data library.\footnote{\url{https://mba.tuck.dartmouth.edu/pages/faculty/ken.french/data_library.html}}

\subsubsection{In-sample and out-of-sample performance}
\label{sec:prediction_performance}

In this section, we leverage the cheap-to-evaluate surrogate to re-estimate the parameters and hidden state of both option pricing models on every day of the sample. We contrast those findings by re-estimating the non-parametric benchmark described in section \ref{sec:rf_presentation} on a daily frequency.

\begin{figure}[t]
\centering     
\includegraphics[width=80mm]{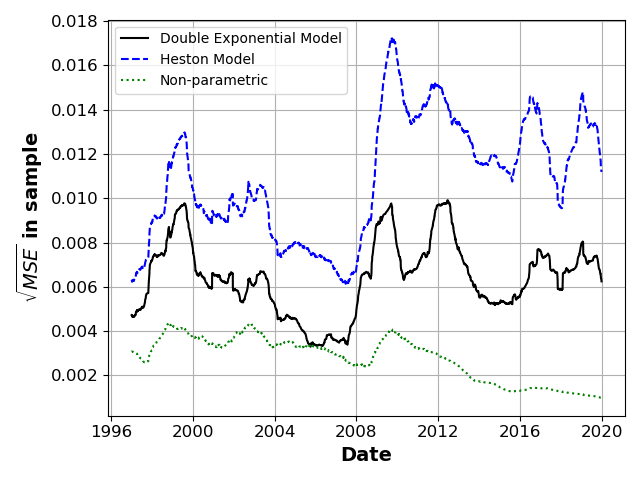}
\includegraphics[width=80mm]{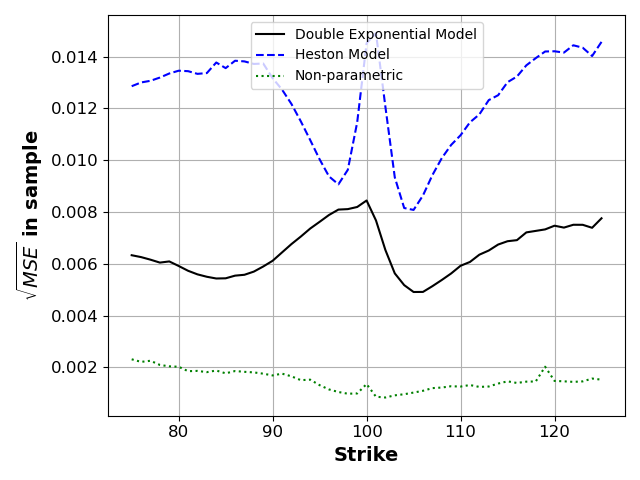}
\caption{The figures above display the mean in-sample performance of the three models across time. We show the mean across days of the $\sqrt{MSE}$. We smooth this measure with a rolling average over the last 252 days.}
\label{fig:first_in_sample}
\end{figure}

We first explore the in-sample performance. In figure \ref{fig:first_in_sample}, we display on the y-axis the daily $\sqrt{MSE}$, smoothed over a rolling average of the previous 252 days.\footnote{To facilitate interpretation of figure~\ref{fig:first_in_sample}, we additionally display the square root of the MSE here as well as throughout the remainder of this section.} As it was to be expected, we find that the parametric model with more degrees of freedom, i.e., the BDJM model, creates a better in-sample fit than the simpler HM. Note that both models' in-sample MSE varies across time, with peaks in 1998 and during the 2008 crisis. We also observe that both parametric models in-sample errors are larger after the 2008 crisis.  Finally, we note that the non-parametric model strongly outperforms both structural models in-sample. 

Next, we turn to the out-of-sample measure of performance. 
Figure~\ref{fig:oos_perf_mean} displays the prediction errors for the two models and the random forest for prediction horizons of 1 to 30 days, $\tau=1,2,...,30$. On the y-axis, we show the mean daily square root of the MSE. The left panel (a) shows the mean daily performance, the middle panel (b) shows the 10th percentile estimated across days, while the right panel (c) shows the 90th percentile. 

On average, the difference in the performance is small for all models. For short prediction horizons $\tau$, the random forest is better than that of the HM. For $\tau=1$, the non-parametric benchmark's performance is almost equal to that of the BDJM, whereas, for longer horizons, the BDJM yields significantly lower prediction errors. Note that the difference in performance between the HM and the BDJM is stable across time.

The percentile graphs reveal that the BDJM  benchmark significantly outperforms both the HM and RF in the 10th percentile measure but not under the 90th percentile. This effect is more substantial for large out-of-sample horizons. This result suggests that the over-performance of the BDJM is concentrated on ``good days''.

\renewcommand{\ww}{0.3\linewidth}
\begin{figure}[t]
\centering     
\subfigure[]{\label{fig:oos_blindmean}\includegraphics[width=\ww]{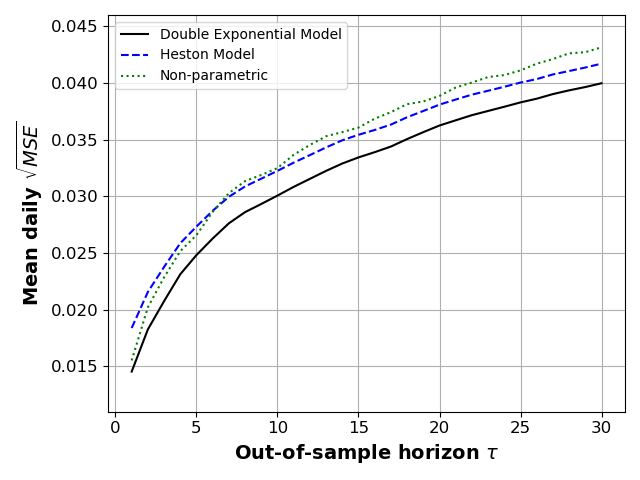}}
\subfigure[]{\label{fig:oos_blindq10}\includegraphics[width=\ww]{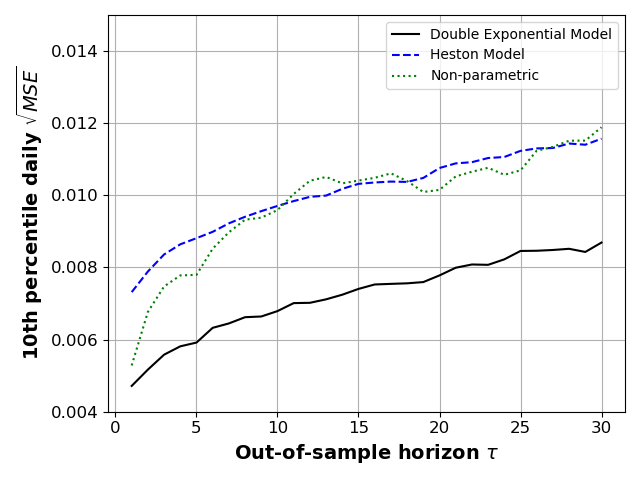}}
\subfigure[]{\label{fig:oos_blindq90}\includegraphics[width=\ww]{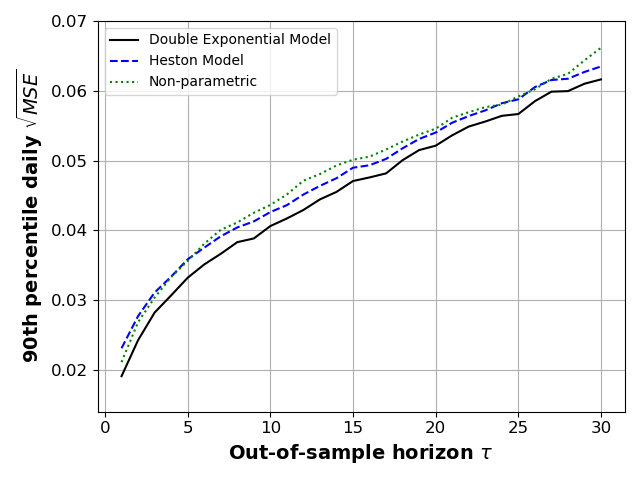}}
\caption{
   The figures above compare the models' out-of-sample performance. On the x-axis, we show the out-of-sample horizon $\tau$, whereas the y-axis depicts the daily pricing error. In the left panel (a), we measure the square root of the daily mean MSE. Panel (b) shows the 10th percentile of the daily mean MSE, while panel (c) shows the 90th percentile.
}
\label{fig:oos_perf_mean}
\end{figure}

Next, we look at the models' performance across time. 
In figure~\ref{fig:oos_ts_blind}, we display the daily $\sqrt{MSE}$ smoothed across time with a rolling average on the last 252 days as a function of time.  Each panel shows the average daily $\sqrt{MSE}$ estimated on a different subsample defined by maturity. In all panels, the out-of-sample performance is measured with a forecasting horizon of one week ($\tau=5$). 

These figures reveal several interesting patterns. First, we note that, while the non-parametric is outperformed on average, the RF does outperform both the BDJM and HM on most days for options with short maturities ($0<T\leq7$ and $7<T\leq30$). Second, the average out of sample errors of all models on options with a long time to maturity is larger after the 2008 crisis. 

We propose two non-exclusive hypotheses to explain the relatively poor performance of parametric models on panels (a) and (b). First, options with small maturities are often less liquid than their counterparts, and while liquidity can be an important pricing factor, it is assumed insignificant by parametric models. Second, most of the academic literature removes short maturity options from sample when testing and discussing models. To do so, they argue that the illiquidity makes these options prices unreliable~\citep[see, e.g.][]{andersen2015parametric,andersen2017short,kadan2020bound}. However, the relatively good performance of our RF shows that there is some reliable pattern in the implied volatility smile.

\renewcommand{\ww}{0.45\linewidth}
\begin{figure}[t]
\centering     
\subfigure[$0<T\leq7$]{\label{fig:a}\includegraphics[width=\ww]{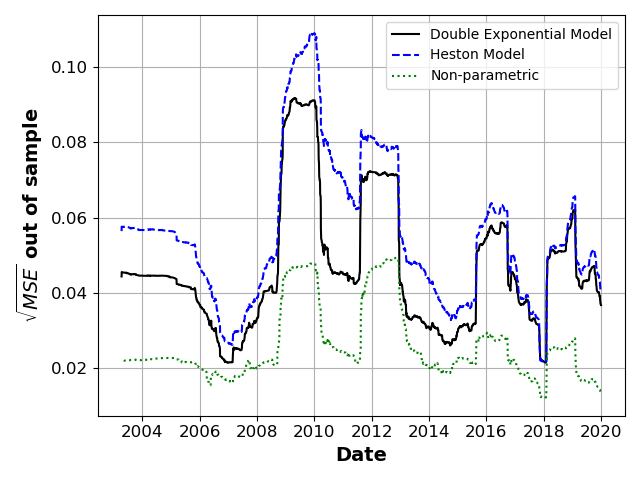}}
\subfigure[$7<T\leq30$]{\label{fig:b}\includegraphics[width=\ww]{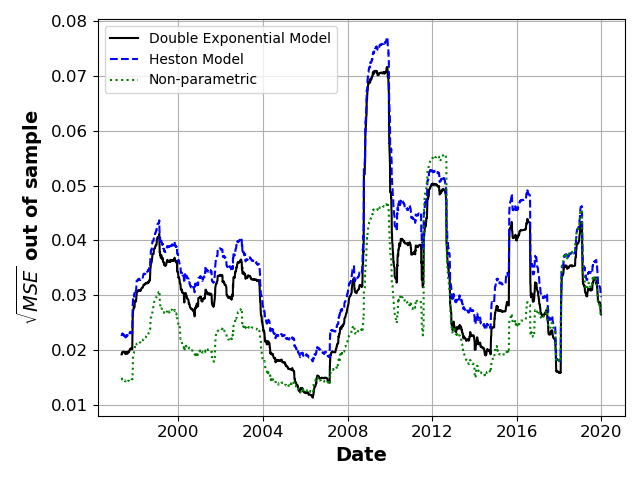}}
\subfigure[$30<T\leq90$]{\label{fig:b}\includegraphics[width=\ww]{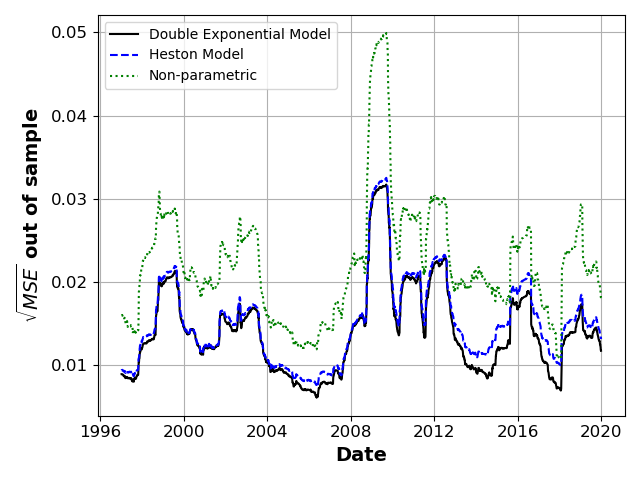}}
\subfigure[$90<T\leq\infty$]{\label{fig:b}\includegraphics[width=\ww]{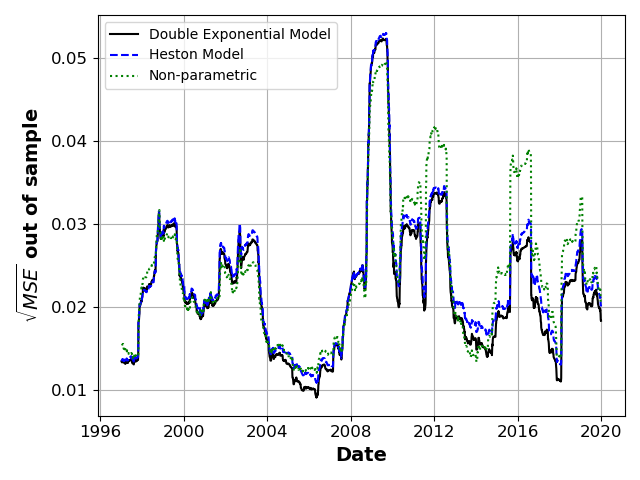}}
\caption{The figures above show the model's out-of-sample performance across time on subsamples defined by maturity brackets. All panels shows the performance for a forecasting horizon of one week ($\tau=5$).
}
\label{fig:oos_ts_blind}
\end{figure}

Figure \ref{fig:oos_cs_blind} further explores the distribution of the out-of-sample performance by showing the average $\sqrt{MSE}$ per standardized strike. As in figure \ref{fig:oos_ts_blind}, we split the sample in each panel by maturity brackets. Panel (b) and (d) show that for some maturity ranges, all models tend to underperform significantly more with deep-out-of-the-money call options than the rest of the sample.\footnote{Recall that the sample does not contain in-the-money options. Therefore, the left-hand side of the graphs ($\hat{K}<100$) contains only put-options.}  On the panel (c), we see that the parametric models significantly outperform the RF for options with a maturity contained between one and three months.  Finally, when looking at the first two panels the relative overperformance of the RF is stable across moneyness on options with a very short time to maturity (a), but concentrated on put options for options with maturities between one week and one month (b).

\renewcommand{\ww}{0.45\linewidth}
\begin{figure}[t]
\centering     
\subfigure[$0<T\leq7$]{\label{fig:a}\includegraphics[width=\ww]{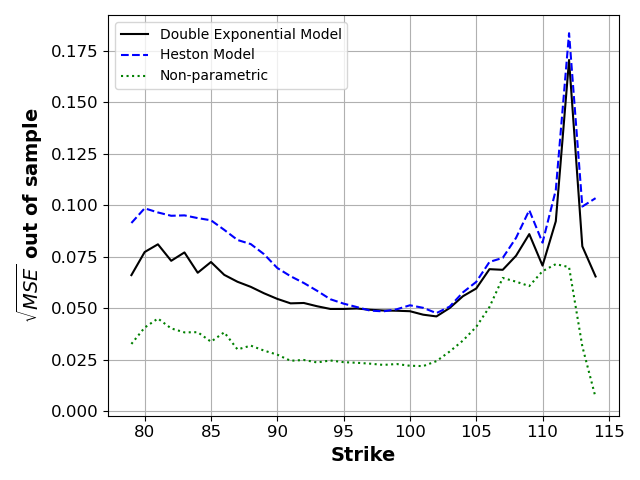}}
\subfigure[$7<T\leq30$]{\label{fig:b}\includegraphics[width=\ww]{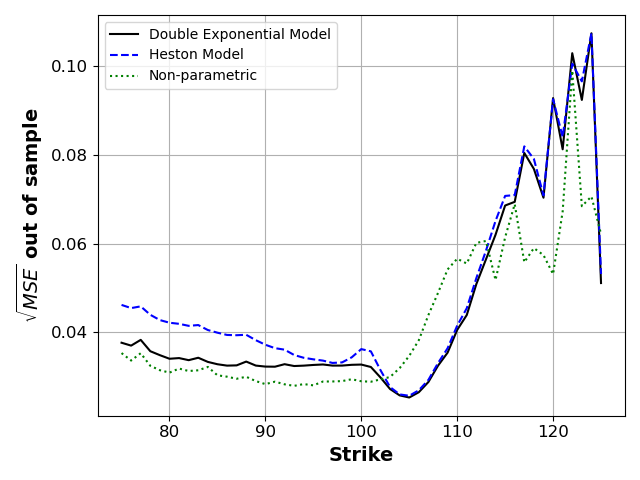}}
\subfigure[$30<T\leq90$]{\label{fig:b}\includegraphics[width=\ww]{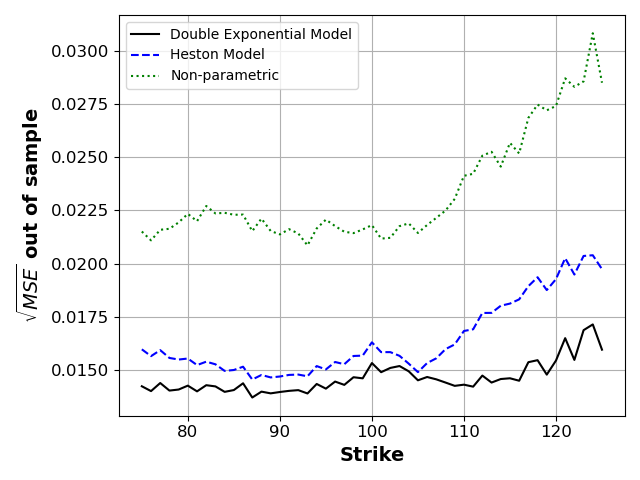}}
\subfigure[$90<T\leq\infty$]{\label{fig:b}\includegraphics[width=\ww]{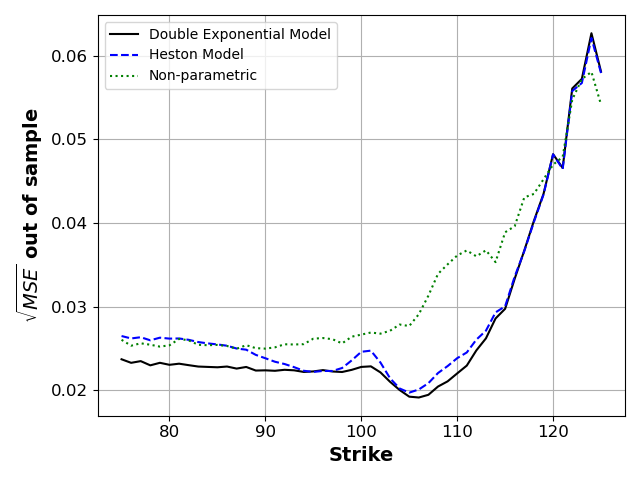}}
\caption{The figures above show the model's out of sample performance across moneyness on subsamples defined by maturity brackets. All panels shows the performance for a forecasting horizon of one week ($\tau=5$)
}
\label{fig:oos_cs_blind}
\end{figure}

Finally, we finish this analysis by looking at the difference in the average daily performance of the models across time. We subtract the daily mean squared error of the HM and RF from that of the BDJM to create two time-series. The more negative values show a larger overperformance of the BDJM. Figure \ref{fig:oos_perf_diff} shows these time-series for different forecasting horizon. 
Comparing the three panels, we see that the volatility of the RF performance increases with the forecasting horizon. At the same time, panels (a) and (b) show that with a small forecasting horizon, the BDJM relative performance from that of the HM significantly increased after the 2008 crisis. This result suggests that before 2008, the market may have omitted to price jump risk properly. After the crisis, however, they either learned or were forced to do so by new legislation.

\renewcommand{\ww}{0.47\linewidth}
\begin{figure}[t]
\centering     
\subfigure[$\tau=1$]{\label{fig:oos_blindmean}\includegraphics[width=\ww]{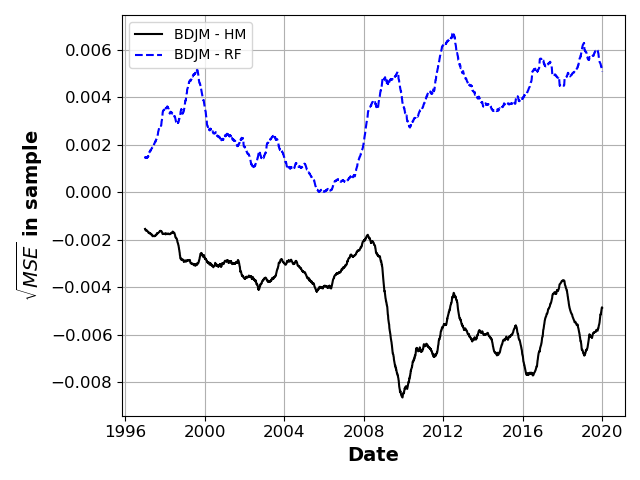}}
\subfigure[$\tau=5$]{\label{fig:oos_blindq10}\includegraphics[width=\ww]{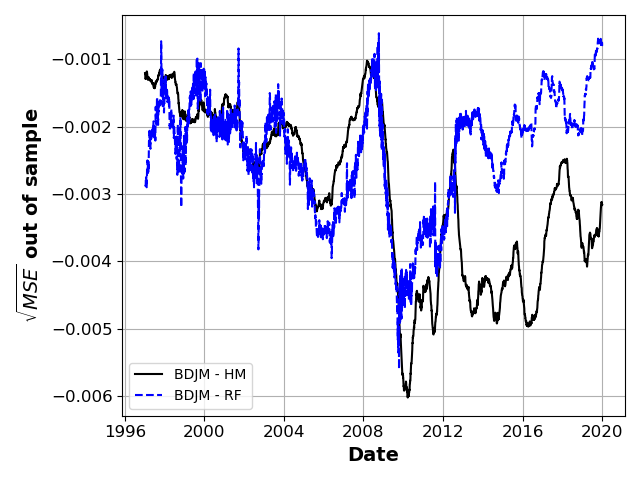}}
\caption{
   The figures above compare the models' out-of-sample performance across time of the HM and RF compared to the best performing model, the BDJM. We first compute the mean daily $\sqrt{MSE}$ for each model, then subtract the performance of the HM and RF to that of the HM to create two time series. Finally, we smooth our results with a 252 day rolling average. The two panels show this measure with a forecasting horizon equal to 1 and 5 business days, respectively.
}
\label{fig:oos_perf_diff}
\end{figure}

Under all specifications, the RM is surprisingly competitive in forecasting BSIVs out-of-sample. The difference in average performance between all models is not large, and the RF does outperform both the BDJM and HM for specific types of options. 

In appendix \ref{sec:additional_analysis}, we complement the analysis discussed in this section with additional figures showing the models' performance on other data subsamples.

\subsubsection{The explaining factors of the out-of-sample performance}

We now investigate what drives the differences in performance among the option pricing models. To do so, we create four variables of interest. The first one is the difference in performance between the daily average MSE of the BDJM and HM $(\sqrt{MSE_{BDJM}}-\sqrt{MSE_{HM}})$. The second one is same difference between BDJM and the non-parametric benchmark, $(\sqrt{MSE_{BDJM}}-\sqrt{MSE_{RF}})$. In addition, we look at the difference in performance for each option $i$ at time $t$, $100*({MSE_{BDJM,i,t}}-{MSE_{HM,i,t}})$ and  $100*({MSE_{BDJM,i,t}}-{MSE_{RF,i,t}})$
For simplicity, we focus on a forecasting horizon of a week ($\tau=5$).

In figure~\ref{fig:auto_corr}, we show the autocorrelations of the time series  $y_t=(\sqrt{MSE_{BDJM}}-\sqrt{MSE_{HM}})$ (a) and $y_t=(\sqrt{MSE_{BDJM}}-\sqrt{MSE_{RF}})$ (b). These graphs show that the difference in performance between the two parametric models has strong and positive autocorrelations. On the other hand, the difference in daily performance between the BDJM and the non-parametric RF has little to no autocorrelation. These results suggest that the state of the economy in which the BDJM is comparatively better suited than the HM is spread across long time periods, while the state of the economy in which the RF performs relatively better than the parametric alternative are short lived.

\renewcommand{\ww}{0.45\linewidth}
\begin{figure}[t]
\centering     
\subfigure[$y_t = (\sqrt{MSE_{BDJM}}-\sqrt{MSE_{HM}})$]{\label{fig:a}\includegraphics[width=\ww]{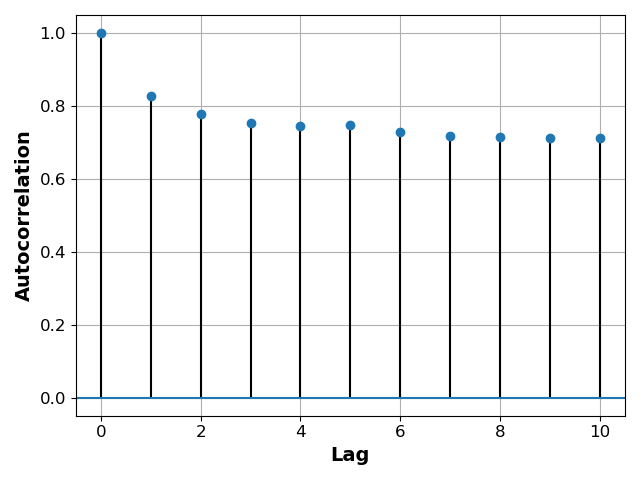}}
\subfigure[$y_t = (\sqrt{MSE_{BDJM}}-\sqrt{MSE_{RF}})$]{\label{fig:b}\includegraphics[width=\ww]{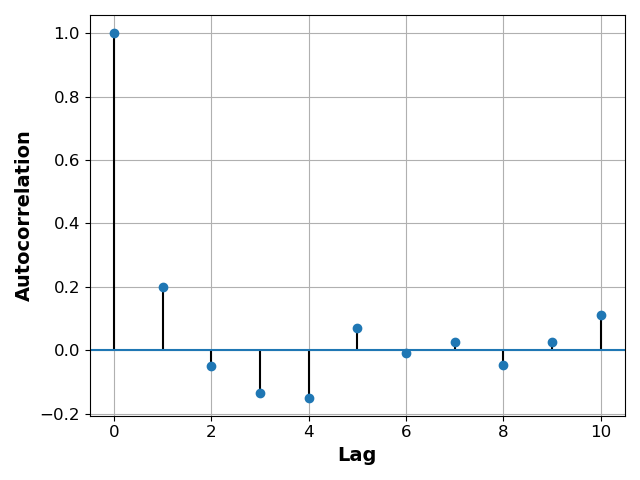}}
\caption{The figures above show the autocorrelations of the relative performance of the BDJM against the HM (a) and RF (b). We compute the difference between the square root of the daily MSE of the models and show the autocorrelations on the y-axis and the lags on the x-axis.
}
\label{fig:auto_corr}
\end{figure}

Next, we investigate potential factors that could explain the relative daily performance among the models. To do so, we compute the mean daily bid-ask spread at time $t$ ($(b-a)_t$) as a proxy of liquidity risk. We capture the relative volatility risk with the VIX index $VIX_t$. We test the risk of jumps in the underlying asset's price with variable $JUMP_t$. Finally, because the jump and volatility risk are highly correlated (0.84), we construct a decorrelated version of the volatility risk. We define $\tilde{VIX}_t$ as the residual from the regression $VIX_t = a + b \cdot JUMP_t + \epsilon_{t}$.

Panel A of table~\ref{table:perf_reg_hbL1} reports various configurations of a regression where the dependent variable is the difference in daily performance between the BDJM and the HM, $y_t = 100(MSE_{BDJM,i,t}-MSE_{HM,i,t})$, and the dependent variable are the market measures of liquidity, volatility, and risks of jumps. All three variables of interest are statistically significant. The liquidity measure's coefficient ($(b-a)_t$) is positive, which means that the BDJM performance relative to that of the HM is worst on days with high liquidity risks. Both parametric models' assumptions include perfect liquidity, and neither is expected to perform well on days where insufficient liquidity has a large impact on price. When faced with an omitted pricing factor, both models misuse their degrees of freedom. Therefore, it is intuitive that the model with more degrees of freedom underperforms more on those low liquidity days. 
The coefficient associated with jump risks is negative, meaning that the BDJM outperforms significantly more on days with large jump risks. This effect is to be expected given that the BDJM was explicitly designed to model such risks. 
Similarly, the BDJM overperform on days with high volatility risks, suggesting that the BDJM is more suited than the HM to predict steep implied volatility smiles.

Panel B of table~\ref{table:perf_reg_hbL1} shows the same regressions when we analyze the difference in performance between the BDJM and RF ($y_t = 100(MSE_{BDJM,i,t}-MSE_{RF,i,t})$). When controlling for the yearly fix effect, the coefficient associated with liquidity risks is positive and statistically significant. These results suggest that the RF relative performance is higher when the liquidity risks are high. This mechanism is coherent because, unlike the RF, the parametric BDJM was specifically designed under the assumption that liquidity had no impact on prices. 
The coefficient associated with jump risk is significant and negative, as it is associated with volatility risks. These two results imply that the BDJM has a small advantage when modeling days with steep implied volatility smile and high jump risks.

\begin{table}[ht!]
   \centering
   \caption{The table below presents various combinations a regression where the dependent variable is, for each option $i$ on day $t$ the difference in performance between the BDJM and the HM ($100(MSE_{BDJM,i,t}-MSE_{HM,i,t})$) in panel A, and the performance difference between the BDHM and RF in panel B ($100(MSE_{BDJM,i,t}-MSE_{RF,i,t})$), and the independent variables include: the bid-ask spread $(b-a)_t$, the volatility index $VIX_t$, the jump risk $JUMP_t$. We estimated model performance with a forecasting horizon of one day $\tau=5$. The values in parentheses are standard errors, while  $^*$ denotes significance at the 10\% level, $^{**}$ at 5\%, and $^{***}$ at 1\%.
}
\resizebox{.99\textwidth}{!}{\begin{tabular}{lcccccccc}
\toprule 
\multicolumn{9}{c}{\parboxc{c}{0.6cm}{Panel A: $y_{i,t} = 100(MSE_{DBJM,i,t}-MSE_{HM,i,t})$}} \\\hline
{} & \parboxc{c}{0.6cm}{(1)} & \parboxc{c}{0.6cm}{(2)} & \parboxc{c}{0.6cm}{(3)} & \parboxc{c}{0.6cm}{(4)} & \parboxc{c}{0.6cm}{(5)} & \parboxc{c}{0.6cm}{(6)} & \parboxc{c}{0.6cm}{(7)} & \parboxc{c}{0.6cm}{(8)} \\
\cmidrule{2-9}
$(b-a)_t$       &                         &                         &      \phantom{*}0.0022* &     \phantom{*}0.0029** &                         &                         &     \phantom{*}0.075*** &    \phantom{*}0.0516*** \\
                &                         &                         &                (0.0012) &     (0.0013)\phantom{*} &                         &                         &     (0.0028)\phantom{*} &    (0.0028)\phantom{**}\smallskip \\
$JUMP_t$        &                         &              -0.0049*** &                         &              -0.0049*** &                         &              -0.0078*** &                         &              -0.0083*** \\
                &                         &    (0.0001)\phantom{**} &                         &    (0.0001)\phantom{**} &                         &    (0.0002)\phantom{**} &                         &    (0.0002)\phantom{**}\smallskip \\
$VIX_t$         &              -0.0005*** &                         &                         &                         &              -0.0006*** &                         &                         &                         \\
                &    (0.0)\phantom{*****} &                         &                         &                         &    (0.0)\phantom{*****} &                         &                         &                        \smallskip \\
$\tilde{VIX_t}$ &                         &                         &                         &              -0.0007*** &                         &                         &                         &              -0.0007*** \\
                &                         &                         &                         &    (0.0)\phantom{*****} &                         &                         &                         &    (0.0)\phantom{*****}\smallskip \\
constant        &              -0.0045*** &              -0.0097*** &              -0.0126*** &                -0.01*** &                         &                         &                         &                         \\
                &    (0.0001)\phantom{**} &    (0.0001)\phantom{**} &    (0.0002)\phantom{**} &                (0.0002) &                         &                         &                         &                         \\
\medskip\\
year FE         &                      No &                      No &                      No &                      No &                     Yes &                     Yes &                     Yes &                     Yes \\
\medskip\\
Observations    &               5,493,397 &               5,484,390 &               5,493,397 &               5,484,390 &               5,493,397 &               5,484,390 &               5,493,397 &               5,484,390 \\
$R^2$           &                  0.0008 &                  0.0003 &                       0.0001 &                   0.001 &                  0.0027 &                  0.0022 &                   0.002 &                  0.0027 \\
\midrule
\\
\multicolumn{9}{c}{\parboxc{c}{0.6cm}{Panel B: $y_{i,t} = 100(MSE_{DBJM,i,t}-MSE_{RF,i,t})$}} \\
\midrule
{} & \parboxc{c}{0.6cm}{(1)} & \parboxc{c}{0.6cm}{(2)} & \parboxc{c}{0.6cm}{(3)} & \parboxc{c}{0.6cm}{(4)} & \parboxc{c}{0.6cm}{(5)} & \parboxc{c}{0.6cm}{(6)} & \parboxc{c}{0.6cm}{(7)} & \parboxc{c}{0.6cm}{(8)} \\
\cmidrule{2-9}
$(b-a)_t$       &                         &                         &              -1.3513*** &               -3.612*** &                         &                         &              15.8412*** &    \phantom{*}7.2956*** \\
                &                         &                         &    (0.5008)\phantom{**} &     (0.5173)\phantom{*} &                         &                         &    (1.1346)\phantom{**} &    (0.7151)\phantom{**}\smallskip \\
$JUMP_t$        &                         &              -0.7315*** &                         &              -0.6506*** &                         &              -0.7505*** &                         &              -1.1787*** \\
                &                         &    (0.0525)\phantom{**} &                         &    (0.0538)\phantom{**} &                         &    (0.0783)\phantom{**} &                         &    (0.0704)\phantom{**}\smallskip \\
$VIX_t$         &              -0.1244*** &                         &                         &                         &              -0.1907*** &                         &                         &                         \\
                &    (0.0028)\phantom{**} &                         &                         &                         &    (0.0039)\phantom{**} &                         &                         &                        \smallskip \\
$\tilde{VIX_t}$ &                         &                         &                         &              -0.2705*** &                         &                         &                         &              -0.3086*** \\
                &                         &                         &                         &    (0.0048)\phantom{**} &                         &                         &                         &    (0.0053)\phantom{**}\smallskip \\
constant        &    \phantom{*}1.1129*** &              -0.5713*** &              -0.7856*** &               -0.1478** &                         &                         &                         &                         \\
                &    (0.0501)\phantom{**} &    (0.0331)\phantom{**} &    (0.067)\phantom{***} &     (0.0691)\phantom{*} &                         &                         &                         &                         \\
\medskip\\
year FE         &                      No &                      No &                      No &                      No &                     Yes &                     Yes &                     Yes &                     Yes \\
\medskip\\
Observations    &               5,523,407 &               5,514,080 &               5,523,407 &               5,514,080 &               5,523,407 &               5,514,080 &               5,523,407 &               5,514,080 \\
$R^2$           &                  0.0004 &                   0.0001 &                   0.0001 &                  0.0006 &                  0.0006 &                  0.0002 &                  0.0002 &                  0.0013 \\
\bottomrule
\end{tabular}
}
\label{table:perf_reg_hbL1}
\end{table}

\subsubsection{Parameter stability}

Our fast-to-evaluate surrogates allow us to swiftly and cheaply create a time-series of daily estimated parameters across time. If one of the option pricing models is correct, we should see only a small variation in the parameters across time, but we measure large variations from one day to the next for both models (c.f., appendix \ref{sec:appendix_par_across_time}). 

We now test statistically whether the models' parameters are stable across time. To do so, we apply the statistical tests developed in section 5.2 by~\cite{andersen2015parametric}. According to their work, given a time-period $t$, and $t+1$, if the pricing model is valid for the two distinct time-periods we have, 

\begin{equation}\label{equ:anderson_test}
\left(\Theta_{t}-\Theta_{t+1}\right)^{\prime}\left(\widehat{\operatorname{Avar}}\left(\Theta_t \right)+\widehat{\operatorname{Avar}}\left(\Theta_{t+1}\right)\right)^{-1}\left(\Theta_{t}-\Theta_{t+1}\right) \stackrel{\mathcal{L}-s}{\longrightarrow} \chi^{2}(q), 
\end{equation}
where $\Theta_{t}$, and $\Theta_{t+1}$ denote the estimated parameters in the time periods $t$ and $t+1$, respectively. $\widehat{\operatorname{Avar}}\left(\Theta_{t}\right)$ and $\widehat{\operatorname{Avar}}\left(\Theta_{t+1}\right)$ denote consistent estimates of the asymptotic variances of these parameters (cf.~\cite{andersen2015parametric}, equations (11)-(12)). 

For every day of the sample $t$, we now estimate the statistical measure of equation \eqref{equ:anderson_test} to test whether the parameters $\Theta_t$ are statistically different from $\Theta_{t+1}$. 

\begin{figure}[t]
\centering     
\subfigure[$\alpha=1\%$]{\label{fig:a}\includegraphics[width=0.6\linewidth]{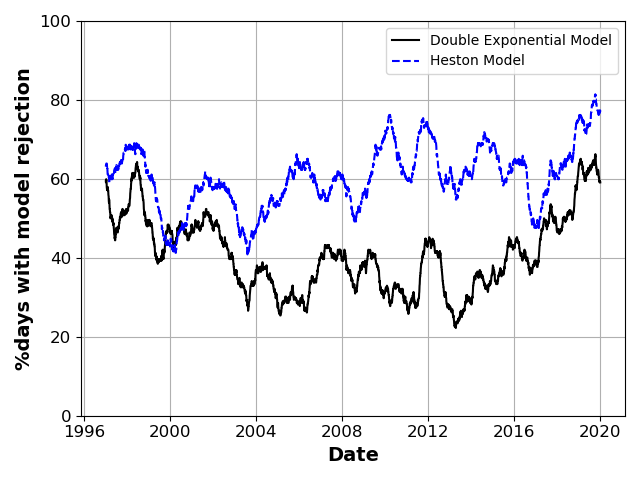}}
\caption{The figure above shows, for each day, the percentage of the last 252 days for which the models' parameters are statistically significantly different from one day to the next. We show the average rejection of the test with a 1\% significance level}
\label{fig:anderson_ts}
\end{figure}

Figure \ref{fig:anderson_ts} shows the results of those daily tests. On the y-axis, we show the average number of days over the past 252 days for which the hypothesis was rejected, that is, the percentage of days for which the models' parameters at time $t$ and $t+1$ are statistically significantly different from one another ($\alpha=1\%$). For all cases, we observe a high rejection rate, which increases after the 2008 crisis. On the whole sample, we reject with 1\% confidence the hypothesis that the models' parameters are stable from one day to the next 41.6\% of the time for the BDJM and 60.7\% of the time for the HM. Those perhaps surprising results suggest that both models' parameters are statistically significantly unstable through time, at least when we estimate the parameters on a daily cross-section of the option. 

\subsubsection{Delta analysis}

Market makers and traders often used models to set-up hedging portfolios, that is, buying and selling other assets to offset or mitigate the risks inherent to holding options. Therefore, the quality of a model's predicted price sensitivity to the underlying assets is often as important as the model's predictive power itself. 

In this section, we compare the BDJM, HM, and non-parametric model's hedging performance. We focus on the \textit{delta}-hedging, that is, the option's price sensitivity to change in the underlying asset. 

Previously, we used the surrogate technology to re-estimate the parameters of the BDJM and HM daily. We now use those daily set of parameters and the original pricing model to estimate for each model $m$ and each option $i$ on day $t$ the first derivative with respect to the underlying asset $\delta_{i,t}^{(m)}$. 

Next, we compute a replication error assuming traders buy the option and counterbalance the \textit{delta}-risk by taking an opposite position on the underlying assets, 
\begin{equation}\label{equ:delta_hedge}
\epsilon^{(m)}_{i,t} = 
\frac{(p_{i,t}-p_{i,t-1}) - \delta_{i,t}^{(m)}(S_t-S_{t-1}) }
{p_{t-1}},
\end{equation}
where $S_t$ is the underlying asset at time $t$, $p_{i,t}$ is the mid quote of the option $i$ at time $t$. We standardize the error by the option price at time $t-1$. Therefore $\epsilon^{(m)}_{i,t}$ represents the replication error in percentage of the original price.\footnote{Unlike the $\sqrt{MSE}$ of the previous section, we standardized the replication errors by the price of the options. We do this standardization here because the replication error is expressed in terms of prices instead of BSIV, and prices vary significantly across time to maturity and moneyness, which can introduce bias in the analysis.}

Table \ref{table:replication_error} shows the mean, standard deviations, and important quantiles of the absolute replications errors across the whole sample ($|\epsilon^{(m)}_{i,t}|$). In addition to the random forest, HM and BDJM, we compute the delta of the Black-Scholes model(BSM).\footnote{For each day, we took the BSIV of at the money options as our estimation of the Black-Scholes implied volatility for all options on that particular day.} A perfect replicating portfolio would produce a replicating error of zero. Note that this is not possible here, as we only hedge the sensitivity to changes in the underlying prices and ignore other well-documented sensitivities: time, volatility, etc. Nonetheless, the relative replicating errors of the various models give a measure of how well suited these models are to hedging applications.

The BDJM produced an average absolute replication error of 0.14175, which only narrowly beats the HM average replication error of 0.15501. The percentile errors suggest that outliers do not produce this over-performance of the BDJM. Finally, while in section \ref{sec:prediction_performance}, we show that the RF is surprisingly competitive when applied to price predictions, table \ref{table:replication_error} shows that the RF is not well suited for hedging applications.\footnote{Table \ref{table:replication_error_large} in appendix \ref{sec:additional_analysis} shows an extended version of this table.} Indeed, the two complex parametric models produced significantly lower replication errors than the RF. 

The relative underperformance of the RF is to be expected. We can view Random forest regressors as a classifier with a large number of small categories. The RF's output is neither continuous nor smooth and is therefore not well suited to produce gradients.

This result highlights a well-known advantage of structural models over non-parametric ones. Non-parametric models are designed to perform specific tasks and need to be adapted for another specific task. On the other hand, a parametric model aims to provide an appropriate description of the world and can therefore be used for multiple applications fairly easily. 

\begin{table}[t!]
   \centering
\caption{The table below shows the mean, standard deviation (std), and important percentiles of the absolute replication error of each model $m$ ($|\epsilon^{(m)}_{i,t}|$) on the whole sample. 
}
{\begin{tabular}{lrrrr}
\toprule
{} &      BSM &       RF &       HM &     BDJM \\
\midrule
mean &  0.19380 &  0.20133 &  0.15501 &  0.14175 \\
std  &  0.42104 &  0.32173 &  0.39458 &  0.28051 \\
5\%   &  0.00423 &  0.00754 &  0.00601 &  0.00595 \\
50\%  &  0.08686 &  0.12512 &  0.07958 &  0.07914 \\
95\%  &  0.65096 &  0.59113 &  0.48294 &  0.44869 \\
\bottomrule
\end{tabular}
}   
\label{table:replication_error}
\end{table}

\begin{figure}[t!]
\centering     
\subfigure[$0<T\leq7$]{\label{fig:delta_ts_T_plot_a}\includegraphics[width=0.45\linewidth]{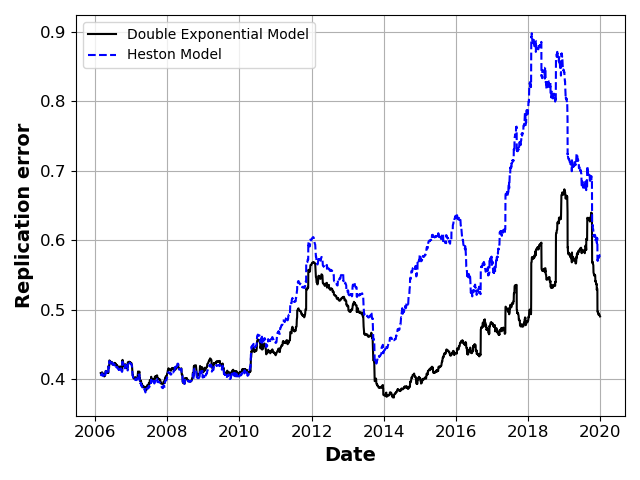}}
\subfigure[$7<T\leq30$]{\label{fig:a}\includegraphics[width=0.45\linewidth]{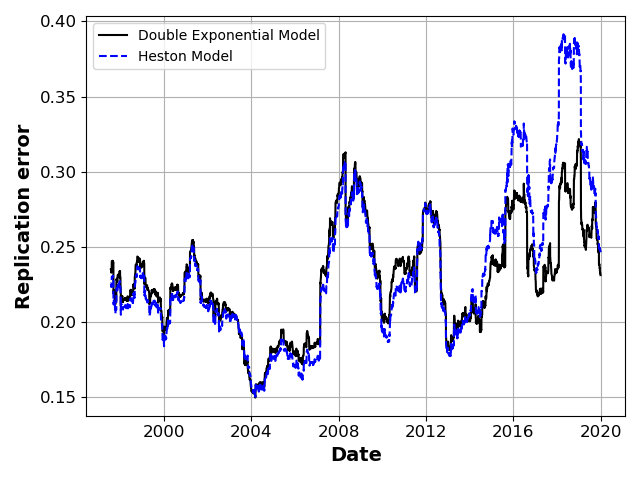}}
\subfigure[$30<T\leq90$]{\label{fig:a}\includegraphics[width=0.45\linewidth]{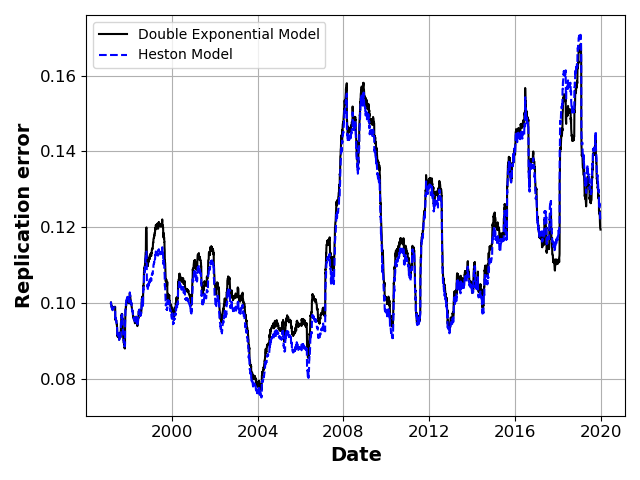}}
\subfigure[$90<T\leq\infty$]{\label{fig:a}\includegraphics[width=0.45\linewidth]{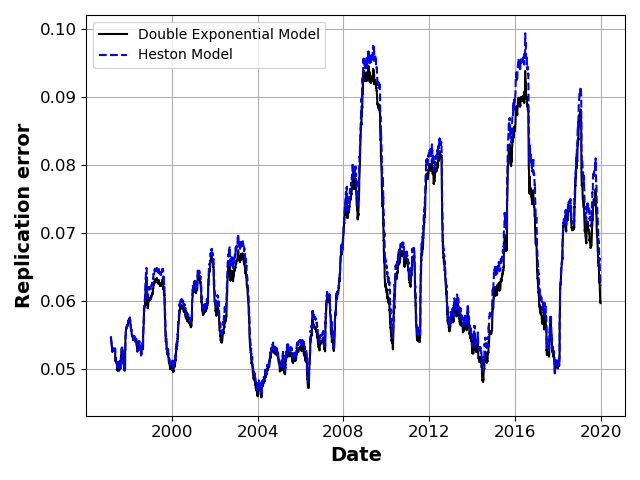}}
\caption{The figures above show the relative mean daily relative replication error smoothed by a 252 days rolling average, on various subsamples defined by the options time to maturity.
}
\label{fig:delta_ts_T_plot}
\end{figure}

We further explore the relationship between replication error, time, and time to maturity  \ref{fig:delta_ts_T_plot}. We show the time series of average daily $\sqrt{MSE}$  on various subsamples defined by time to maturity. The first two panel show options with a relatively low time to maturity ($0<T\leq7$, and ($7<T\leq30$) while the last two shows the performance on the subsample of options with a relatively long time to maturity ($30<T\leq90$, and $90<T\leq\infty$). 

These graphs highlight several interesting patterns. First, we see again that most of the BDJM improvement on the HM concerns short term options. This observation is in line with 
\cite{andersen2017short} who highlight that weekly options are very sensitive to jump risks. Second, we see that the upward trends in replication errors after the 2008 crisis exists across different time to maturity subsample and can therefore not be explained by the change in market composition highlighted in figure \ref{fig:daily_sample_mat}. Finally, figure \ref{fig:delta_ts_T_plot_a} shows that the difference in hedging performance between the two models on options with a short time to maturity does not exist in the earlier part of the sample and only starts to appear around 2011.

\begin{figure}[t!]
\centering     
\subfigure[$0<T\leq7$]{\label{fig:delta_ts_T_plot_a}\includegraphics[width=0.45\linewidth]{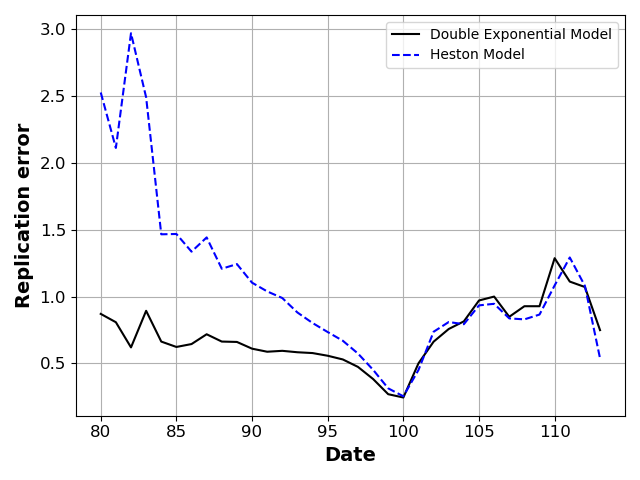}}
\subfigure[$7<T\leq30$]{\label{fig:a}\includegraphics[width=0.45\linewidth]{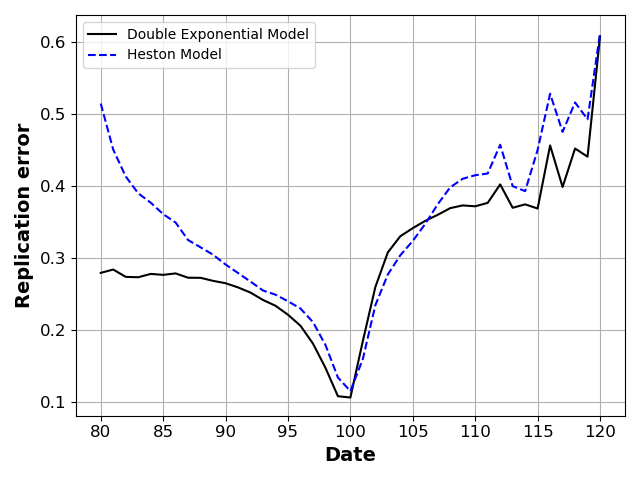}}
\subfigure[$30<T\leq90$]{\label{fig:a}\includegraphics[width=0.45\linewidth]{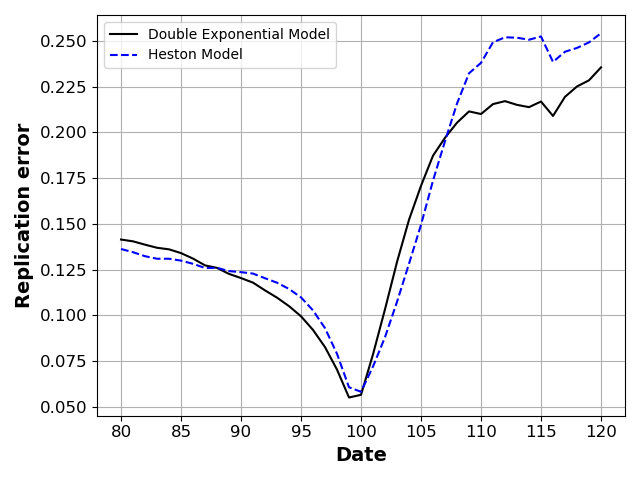}}
\subfigure[$90<T\leq\infty$]{\label{fig:a}\includegraphics[width=0.45\linewidth]{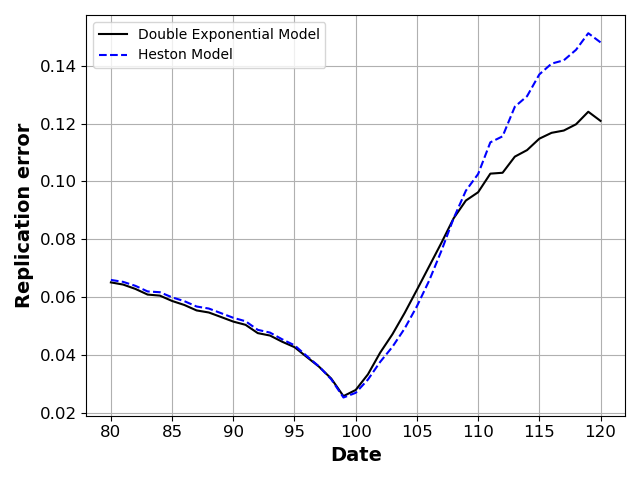}}
\caption{The figures above show the relative mean daily relative replication error across standardized strikes on various subsamples defined by the options time to maturity.
}
\label{fig:delta_cs_T_plot}
\end{figure}

Next, we discuss the relationship between replication error, time, and moneyness with figure \ref{fig:delta_cs_T_plot}. We show the average $\sqrt{MSE}$ per strike level on various subsample defined by time to maturity. 

These graphs show that the replication error is smaller for at-the-money options across all maturity brackets. Indeed, a smile-like shape, reminiscent of the implied volatility smile, appears across all moneyness. Furthermore, panel (c) and (d) show that for options with a maturity larger or equal to 1 month, the replication error is larger for the call options in the sample than the put options. Finally, we see that the overperformance of the BDJM is mostly caused by the hedging of the out-of-the-money put options.

To conclude this analysis of hedging performance, we perform the following regression: 

We use the model's states and parameters we estimated in the previous section to compute the theoretical $\delta$ of each option of the subsamples and compute the following regression,
\begin{equation}\label{equ:delta_reg}
\begin{array}{rcl} 
p_{i, t+1}-p_{i, t} = \beta_{M} \delta_{i,M} (S_{t+1}-S_{t}) &+& \\
\beta_T (T_{i,t+1}-T_{i,t})+
\beta_{J}  (-\mathbbm{1}_{put}) (JUMP_{t+1}-JUMP_{t}) &+& \\\beta_{V} (VIX_{t+1}-VIX_{t}),
\end{array}
\end{equation}
where $p_{i, t+1}-p_{i, t}$ is the change in price of an option $i$ from time $t$ to time $t+1$. $\delta_{i,M}$ is the theoretical delta of the option $i$, predicted by the model $M$. $(S_{t+1}-S_{t})$ is the change in price of the underlying asset. $(T_{i,t+1}-T_{i,t})$ is the change in maturity in days of the asset.\footnote{Because $t$ denotes trading days and not calendar days, the change in time to maturity of an option from time $t$ to $t+1$ is not always equal to 1. For example, after a Friday, two calendar days separate two trading days.} $JUMP_t-JUMP_{t-1})(-\mathbbm{1}_{call})$ denotes the change in the estimated jump risk. We multiply the change in jump risk by -1 for call options to account for the fact that jump risks affect put and call options differently. Finally, and $(VIX_{t+1}-VIX_{t})$ is the change in the VIX index which captures the change in the volatility premium.

We present the estimation of this regression's configurations in table \ref{table:delta_reg}. The $\beta_{M}$ of the HM and BDJM are not equal to 1 in any such configuration. However, the values get closer to 1 as we add controls. Finally, we note that the BDJM-\textit{Delta} produces a higher adjusted $R^2$ than the HM-\textit{Delta}.  

\begin{table}[t!]
  \centering
     \caption{This table shows the estimations of various configurations of equation \eqref{equ:delta_reg}. The values in parentheses are standard errors, while  $^*$ denotes significance at the 10\% level, $^{**}$ at 5\%, and $^{***}$ at 1\%
}
\resizebox{.99\textwidth}{!}{\begin{tabular}{lcccccccc}
\toprule
{} & \parboxc{c}{0.6cm}{(1)} & \parboxc{c}{0.6cm}{(2)} & \parboxc{c}{0.6cm}{(3)} & \parboxc{c}{0.6cm}{(4)} & \parboxc{c}{0.6cm}{(5)} & \parboxc{c}{0.6cm}{(6)} & \parboxc{c}{0.6cm}{(7)} & \parboxc{c}{0.6cm}{(8)} \\
\midrule
$T_t-T_{t-1}$                              &                         &                         &                         &                         &              -0.2178*** &              -0.1453*** &              -0.1561*** &              -0.1566*** \\
                                           &                         &                         &                         &                         &    (0.0008)\phantom{**} &    (0.0004)\phantom{**} &    (0.0004)\phantom{**} &    (0.0004)\phantom{**}\smallskip \\
($JUMP_t-JUMP_{t-1})(-\mathbbm{1}_{call})$ &                         &                         &                         &                         &               10.101*** &    \phantom{*}2.8129*** &    \phantom{*}1.4542*** &    \phantom{*}1.3527*** \\
                                           &                         &                         &                         &                         &     (0.0664)\phantom{*} &    (0.029)\phantom{***} &    (0.0346)\phantom{**} &    (0.0331)\phantom{**}\smallskip \\
$VIX_t-VIX_{t-1}$                          &                         &                         &                         &                         &    \phantom{*}0.8189*** &    \phantom{*}0.9727*** &    \phantom{*}1.0647*** &    \phantom{*}1.0683*** \\
                                           &                         &                         &                         &                         &    (0.0009)\phantom{**} &    (0.0004)\phantom{**} &    (0.0005)\phantom{**} &    (0.0005)\phantom{**}\smallskip \\
$\delta_{i,t}^{(BDJM)}(S_t-S_{t-1})$       &                         &                         &                         &     \phantom{*}0.879*** &                         &                         &                         &    \phantom{*}0.9158*** \\
                                           &                         &                         &                         &     (0.0003)\phantom{*} &                         &                         &                         &    (0.0002)\phantom{**}\smallskip \\
$\delta_{BS}(S_t-S_{t-1})$                 &                         &      \phantom{*}0.92*** &                         &                         &                         &    \phantom{*}0.9361*** &                         &                         \\
                                           &                         &                (0.0002) &                         &                         &                         &    (0.0002)\phantom{**} &                         &                        \smallskip \\
$\delta_{i,t}^{(hm)}(S_t-S_{t-1})$         &                         &                         &    \phantom{*}0.9129*** &                         &                         &                         &    \phantom{*}0.9509*** &                         \\
                                           &                         &                         &    (0.0003)\phantom{**} &                         &                         &                         &    (0.0002)\phantom{**} &                        \smallskip \\
$\delta_{i,t}^{(RF)}(S_t-S_{t-1})$         &              82.5215*** &                         &                         &                         &              80.4929*** &                         &                         &                         \\
                                           &    (0.0603)\phantom{**} &                         &                         &                         &    (0.056)\phantom{***} &                         &                         &                         \\
\medskip\\
Observations                               &               5,021,299 &               5,021,299 &               5,021,299 &               5,021,299 &               5,021,299 &               5,021,299 &               5,021,299 &               5,021,299 \\
$R^2$                                      &                  0.2717 &                  0.7455 &                  0.6702 &                  0.6829 &                  0.3768 &                  0.8819 &                   0.832 &                  0.8457 \\
\bottomrule
\end{tabular}
}

\label{table:delta_reg}
\end{table}

\section{Conclusion}
In this paper, we introduce \textit{deep structural estimation}: a generic framework to swiftly estimate complex structural models in economics and finance. We treat the models' parameters as pseudo-state variables to create a deep-neural network surrogate for replicating the target model. By alleviating the curse of dimensionality, this surrogate approach considerably lowers the computational cost for the prediction and parameter estimation on data. Such a speed gain is non-trivial, as high computational costs often prohibit important analysis of structural models, including a) out-of-sample analysis, b) testing of parameters stability and, c) re-calibration and testing on multiple subsamples. All three of these examples require a fast re-estimation of structural models, which is often infeasible without the surrogate technology. 

We illustrate the validity, performance, and usefulness of the introduced method in the context of financial models by constructing surrogates for two well-known option pricing models: the HM and the BDJM.

First, we demonstrate, with the aid of simulated data, the performance of the surrogates models. We show a) that the surrogate is capable of approximating a highly complex volatility surface with virtually no pricing error, and b) that the surrogate can be used to estimate the hidden states and parameters from a small cross-section of simulated option prices. 

Second, we apply our \textit{deep structural estimation framework} to re-estimate both option pricing models' parameters and hidden states on every trading day of the last 17 years on the cross-section S\&P500 options. We then conduct a thorough in and out-of-sample performance analysis of the structural models. This exercise highlights several interesting patterns: a) while the BDJM outperforms the HM and RF on average, the non-parametric model is more efficient on specific areas of the volatility curve, namely options with a maturity of less than one week and put-options with a maturity of less than one month, b) when comparing the model's hedging performance out-of-sample we show that while the BDJM outperforms the RF, this difference is concentrated on very short maturity options and only occurs after 2011, c) the parameters of the BDJM and HM are relatively unstable through time as we can reject the hypothesis that the parameters do not change from one day to the next 41.6\% of the time for the BDJM and 60.7\% of the time for the HM. 

These results help to identify the strength and weaknesses of the current option pricing theory. On the one hand, the fact that, on average, the BDJM outperforms the RF, whereas HM fails to do so for short forecasting horizons, implies that the extension of the HM to include jump risks was important and that the more modern model reflects market realities better. On the other hand, the poor parameter stability of HM and BDJM and high pricing errors of these models on options with short time to maturity: a) highlights the need for further progress, and b) suggests avenues and direction where the said progress should be directed. Taken together, these insights showcase the usefulness of the \textit{deep structural estimation} framework and the thorough out-of-sample analysis it allows.

\label{sec:Conclusion}

\clearpage

\appendix

\section{Simulated results}
\label{sec:appendix_simulated}
 
 \begin{table}[ht!]
   \centering
     \caption{This table presents the ranges for the surrogate model of the  HM's training sample.}
{\begin{tabular}{lrr}
\toprule
{} &  $\underline{x}^{(j)}$ &  $\bar{x}^{(j)}$ \\
$j$      &                        &                  \\
\midrule
m        &                  -9.00 &            5.000 \\
$rf$     &                   0.00 &            0.075 \\
dividend &                   0.00 &            0.050 \\
$v_t$    &                   0.01 &            0.900 \\
$T$      &                   1.00 &          365.000 \\
$\kappa$ &                   0.10 &           50.000 \\
$\theta$ &                   0.01 &            0.900 \\
$\sigma$ &                   0.10 &            5.000 \\
$\rho$   &                  -1.00 &           -0.000 \\
\bottomrule
\end{tabular}
}
\label{table:hm_range}
\end{table}
\begin{table}[ht!]
   \centering
   \caption{This table presents the ranges for the surrogate model of the  BDJM's training sample.}   
{\begin{tabular}{lrr}
\toprule
{} &  $\underline{x}^{(j)}$ &  $\bar{x}^{(j)}$ \\
$j$       &                        &                  \\
\midrule
m         &                  -9.00 &            5.000 \\
$rf$      &                   0.00 &            0.075 \\
dividend  &                   0.00 &            0.050 \\
$v_t$     &                   0.01 &            0.900 \\
$T$       &                   1.00 &          365.000 \\
$\kappa$  &                   0.10 &           50.000 \\
$\theta$  &                   0.01 &            0.900 \\
$\sigma$  &                   0.10 &            5.000 \\
$\rho$    &                  -1.00 &           -0.000 \\
$\lambda$ &                   0.00 &            4.000 \\
$\nu_1$   &                   0.00 &            0.400 \\
$\nu_2$   &                   0.00 &            0.400 \\
p         &                   0.00 &            1.000 \\
\bottomrule
\end{tabular}
}
\label{table:bdjm_range}
\end{table}

%
\begin{figure}[ht!]
\centering     
\subfigure[]{\label{fig:hest_call_pred_err}\includegraphics[width=70mm]{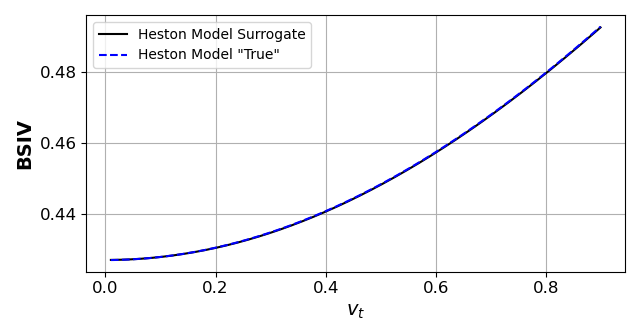}}
\subfigure[]{\label{fig:hest_call_pred_err}\includegraphics[width=70mm]{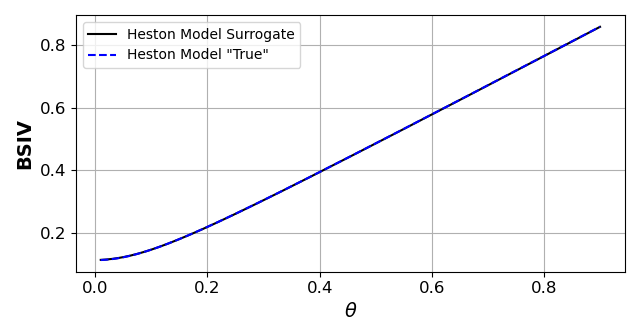}}
\subfigure[]{\label{fig:hest_call_pred_err}\includegraphics[width=70mm]{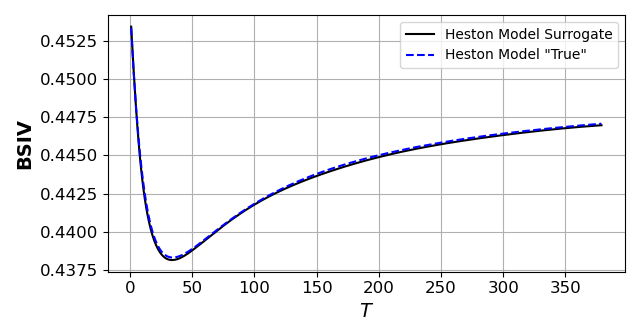}}
\subfigure[]{\label{fig:hest_call_pred_err}\includegraphics[width=70mm]{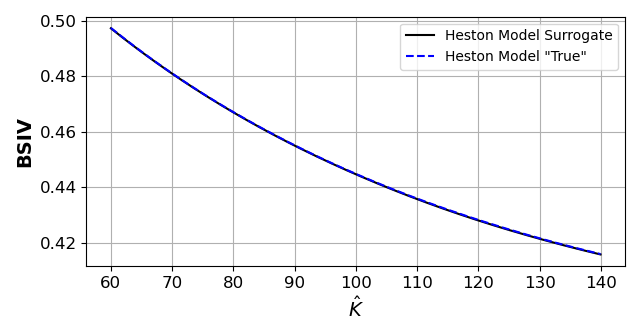}}
\subfigure[]{\label{fig:hest_call_pred_err}\includegraphics[width=70mm]{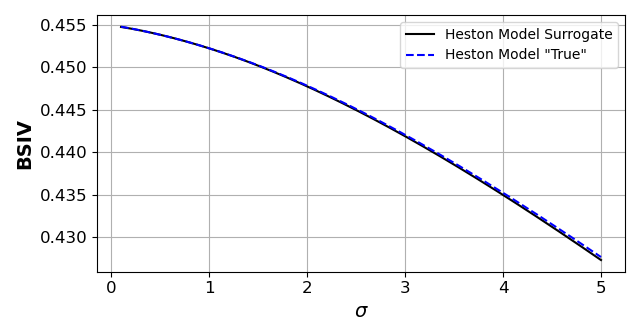}}
\subfigure[]{\label{fig:hest_call_pred_err}\includegraphics[width=70mm]{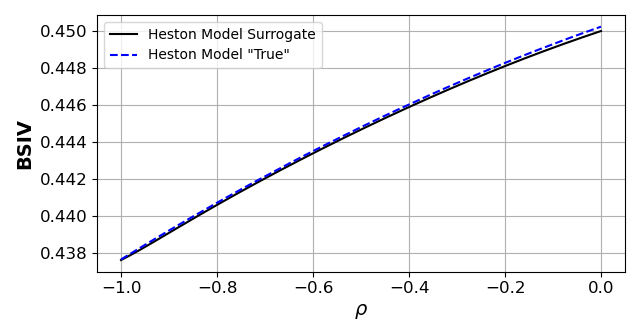}}
\subfigure[]{\label{fig:hest_call_pred_err}\includegraphics[width=70mm]{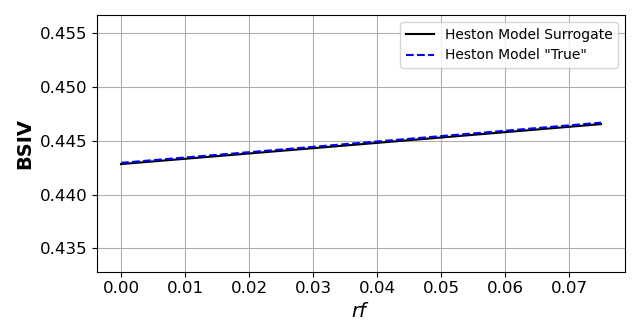}}
\subfigure[]{\label{fig:hest_call_pred_err}\includegraphics[width=70mm]{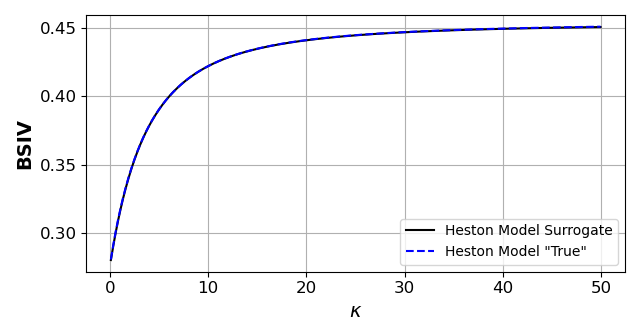}}
\caption{The figures above compare the BSIVs of the surrogate against the predictions of the ``true" HM. We populate the state space with points, where we keep all parameters and states fixed at the mid-range of possible values: $\hat{K}=100.0,rf=0.0375, T=190, \kappa=25.05, \theta=0.455, v_t=0.455, \sigma=2.55$. On each panel, we show the BSIV predicted by the model and its surrogate while varying one of the parameters or states in its admissible range (cf. table~\ref{table:hm_range}).}
\label{fig:sensitivity_hm}
\end{figure}

\begin{figure}[ht!]
\centering     
\subfigure[]{\label{fig:hest_call_pred_err}\includegraphics[width=70mm]{fig/sens/bate_call_good/v0.png}}
\subfigure[]{\label{fig:hest_call_pred_err}\includegraphics[width=70mm]{fig/sens/bate_call_good/theta.png}}
\subfigure[]{\label{fig:hest_call_pred_err}\includegraphics[width=70mm]{fig/sens/bate_call_good/T.png}}
\subfigure[]{\label{fig:hest_call_pred_err}\includegraphics[width=70mm]{fig/sens/bate_call_good/strike.png}}
\subfigure[]{\label{fig:hest_call_pred_err}\includegraphics[width=70mm]{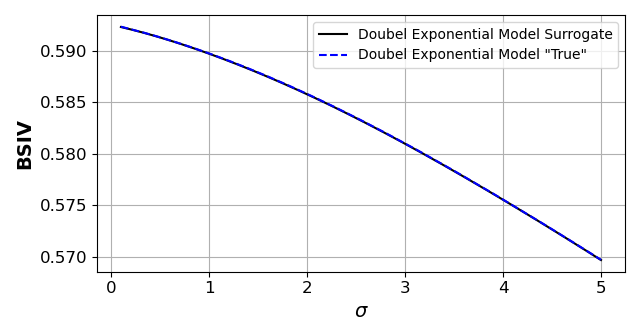}}
\subfigure[]{\label{fig:hest_call_pred_err}\includegraphics[width=70mm]{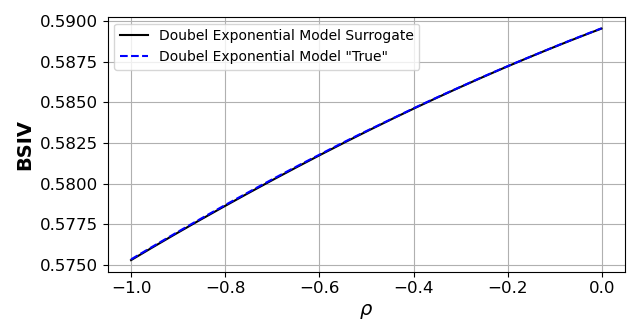}}
\subfigure[]{\label{fig:hest_call_pred_err}\includegraphics[width=70mm]{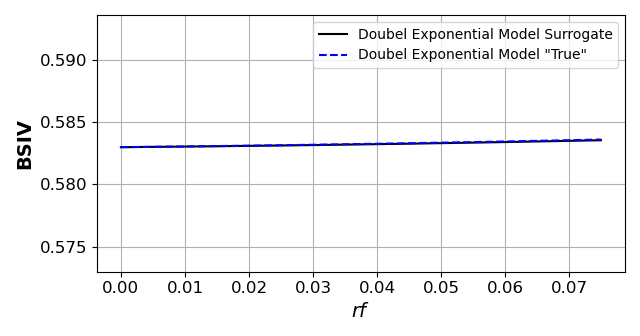}}
\subfigure[]{\label{fig:hest_call_pred_err}\includegraphics[width=70mm]{fig/sens/bate_call_good/kappa.png}}
\caption{The figures above compare the theoretical BSIVs predicted by the surrogate model of the BDJM as done in figure \ref{fig:sensitivity_hm}. The panel above shows sensitivity graphs for all states and parameters in common between the BDJM and HM.}
\label{fig:sensitivity_bate_1}
\end{figure}

\begin{figure}[ht!]
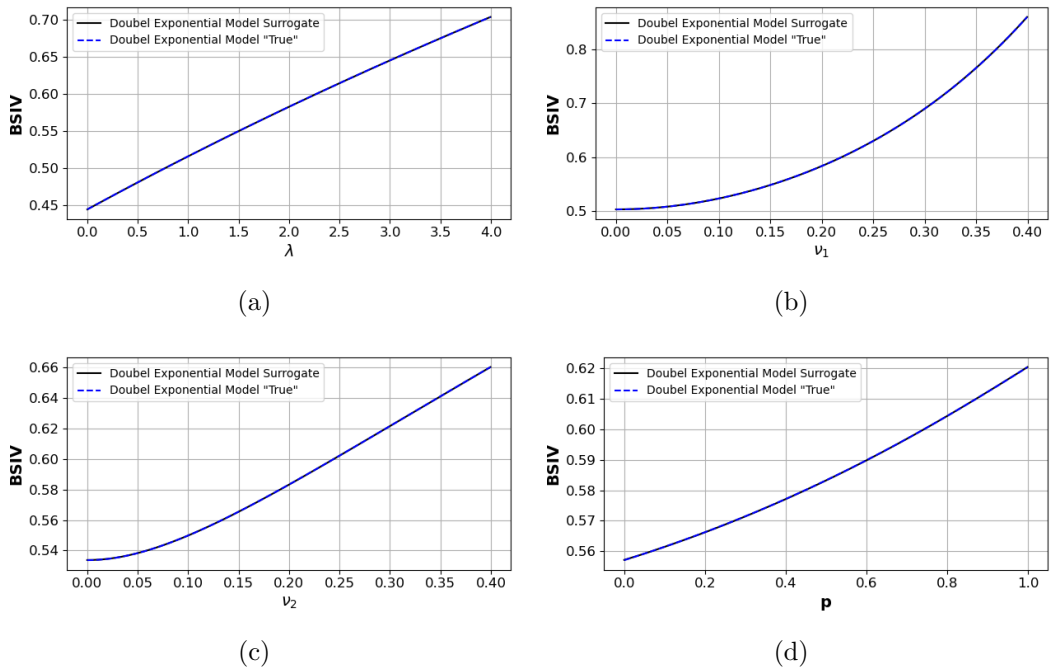

\centering     
\subfigure[]{\label{fig:hest_call_pred_err}\includegraphics[width=70mm]{fig/sens/bate_call_good/lambda_parameter.png}}
\subfigure[]{\label{fig:hest_call_pred_err}\includegraphics[width=70mm]{fig/sens/bate_call_good/nuUp.png}}
\subfigure[]{\label{fig:hest_call_pred_err}\includegraphics[width=70mm]{fig/sens/bate_call_good/nuDown.png}}
\subfigure[]{\label{fig:hest_call_pred_err}\includegraphics[width=70mm]{fig/sens/bate_call_good/p.png}}
\caption{The figures above complement the sensitivity analysis presented in figure \ref{fig:sensitivity_bate_1}. The panel above shows sensitivity graphs for all states and parameters unique to the BDJM model.}
\label{fig:sensitivity_bate_2}
\end{figure}

\begin{figure}[ht!]
\centering     
\subfigure[]{\label{fig:hest_call_pred_err}\includegraphics[width=70mm]{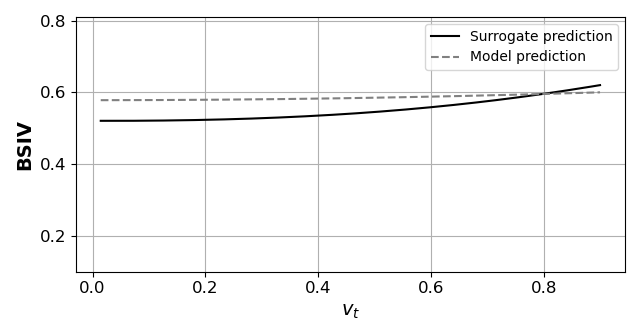}}
\subfigure[]{\label{fig:hest_call_pred_err}\includegraphics[width=70mm]{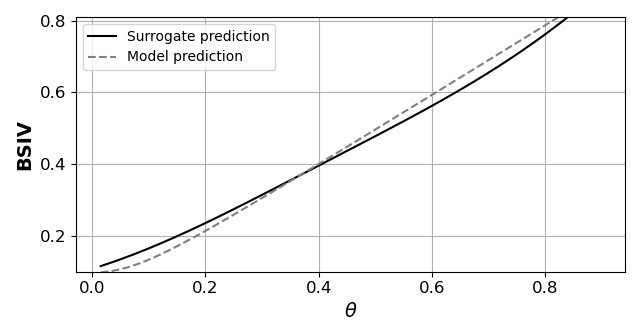}}
\subfigure[]{\label{fig:hest_call_pred_err}\includegraphics[width=70mm]{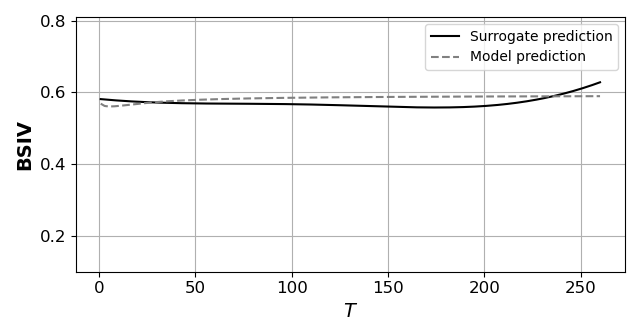}}
\subfigure[]{\label{fig:hest_call_pred_err}\includegraphics[width=70mm]{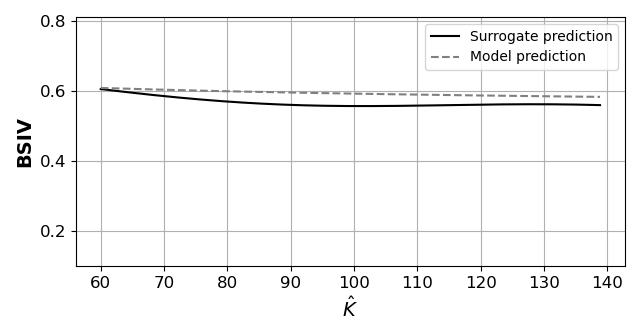}}
\subfigure[]{\label{fig:hest_call_pred_err}\includegraphics[width=70mm]{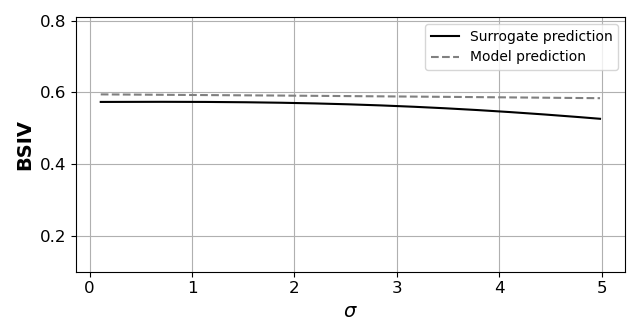}}
\subfigure[]{\label{fig:hest_call_pred_err}\includegraphics[width=70mm]{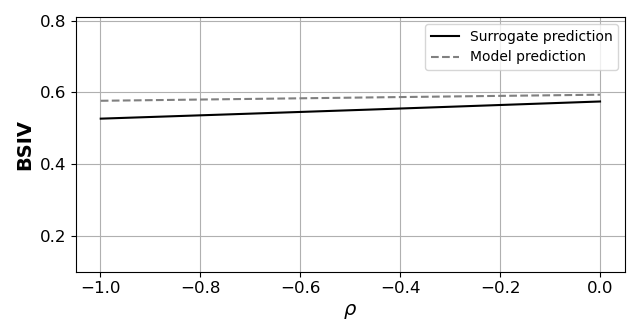}}
\subfigure[]{\label{fig:hest_call_pred_err}\includegraphics[width=70mm]{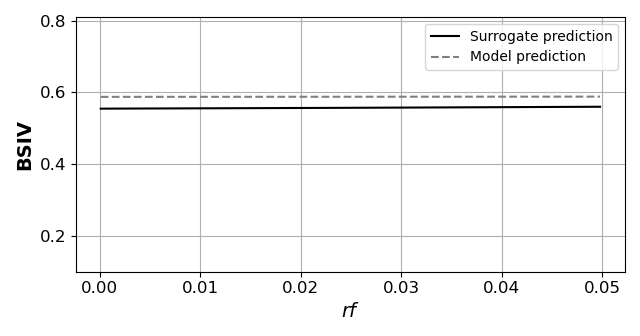}}
\subfigure[]{\label{fig:hest_call_pred_err}\includegraphics[width=70mm]{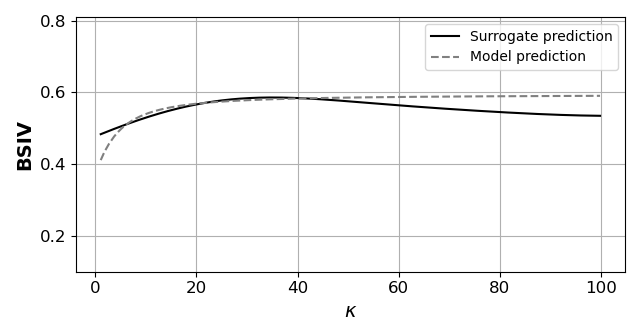}}
\caption{The figures above show results similar to those displayed in figure \ref{fig:sensitivity_hm}. However, we computed here the surrogate's prediction with a shallow neural network that is composed of only one layer with 400 neurons and \textit{Swish} activations functions.}
\label{fig:sensitivity_shallow}
\end{figure}

\begin{figure}[ht!]
\centering     
\includegraphics[width=80mm]{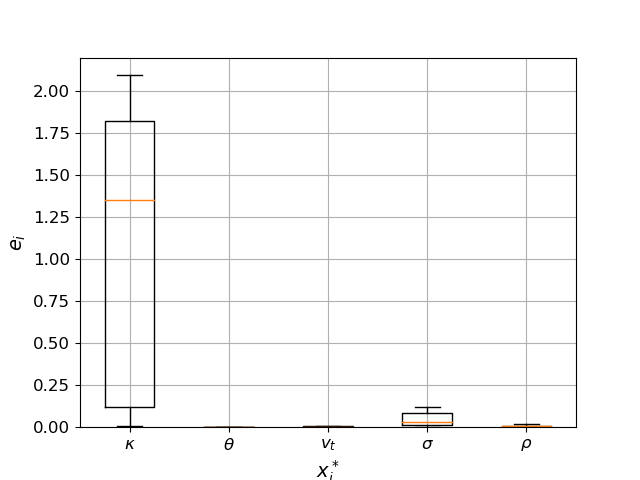}
\caption{The figure above shows the prediction error for the call model of the HM using a shallow network (1 layer of 400 neurons).}.
\label{fig:error_plot_shallow}
\end{figure}

\clearpage

\section{Parameters across time}
\label{sec:appendix_par_across_time}
\begin{figure}[ht!]
\centering     
\includegraphics[width=80mm]{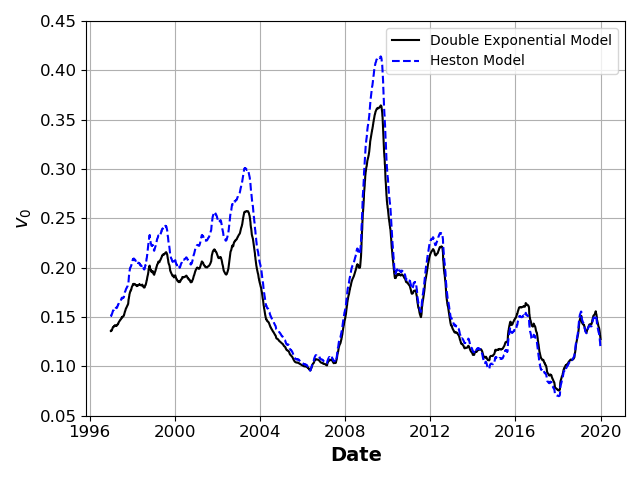}
 \caption{
    The figure above depicts the value of the hidden volatility state $v_t$ estimated every day along with each model's parameter. We smoothed the values with a rolling mean on the last 252 days.
}
\label{fig:time_serie_vola}
\end{figure}

Figure~\ref{fig:time_serie_vola} shows the time series of the hidden state $v_t$ re-estimated at a daily frequency and smooth with a rolling average over the last 252 days. The left panel (a) shows the value estimated with the HM, and the right panel (b) depicts the value estimated with the BDJM panel. We can see that both time-series are highly correlated across time. Furthermore, the volatility estimated with the HM is slightly higher than that estimated by the BDJM. 

Figures~\ref{fig:time_serie_params_BDJM} and \ref{fig:time_serie_params_HM} in appendix \ref{sec:appendix_par_across_time} display the value of the parameters re-estimated daily for both models. We smooth each parameter across time with a rolling average over the last 252 days. We can see a strong variation of all parameters across time. 

\begin{figure}[ht!]
\centering     
\subfigure[$\kappa$]{\label{fig:a}\includegraphics[width=60mm]{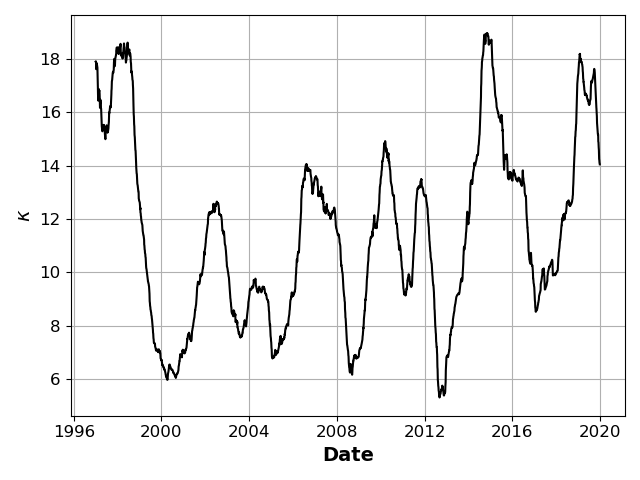}}
\subfigure[$\rho$]{\label{fig:a}\includegraphics[width=60mm]{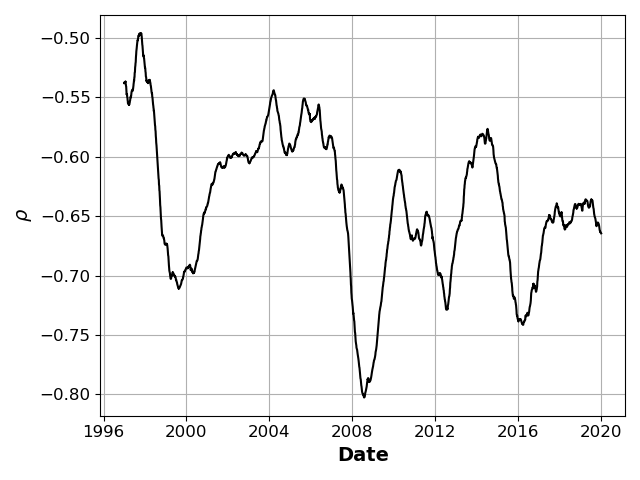}}
\subfigure[$\sigma$]{\label{fig:a}\includegraphics[width=60mm]{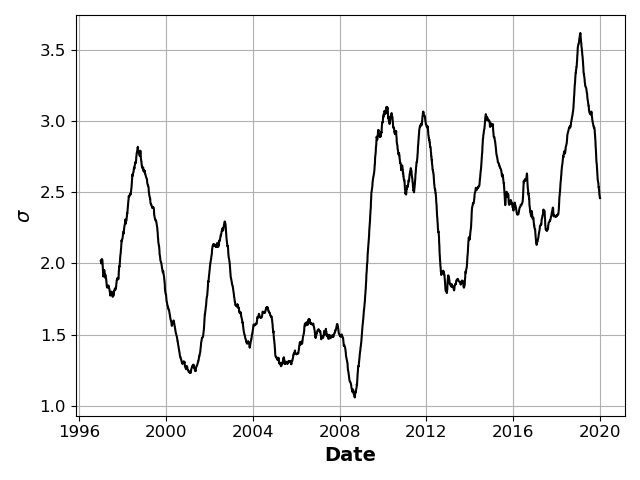}}
\subfigure[$\theta$]{\label{fig:a}\includegraphics[width=60mm]{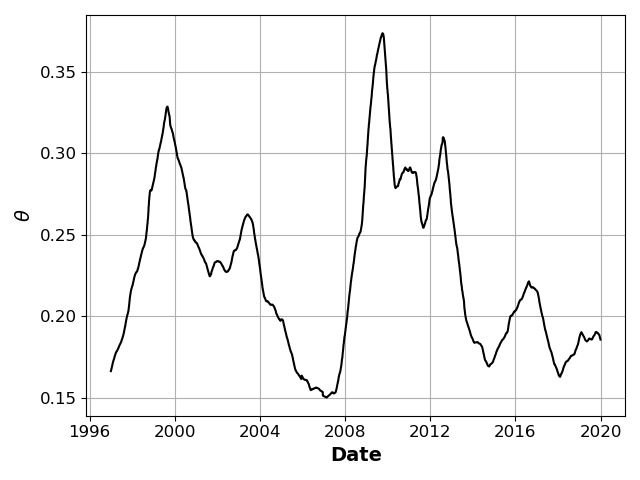}}
    \caption{
    The figures above show the daily estimation parameters of the HM. We smoothed the values with a rolling mean on the last 252 days. 
}
\label{fig:time_serie_params_HM}
\end{figure}

\begin{figure}[ht!]
\centering     
\subfigure[$\kappa$]{\label{fig:a}\includegraphics[width=60mm]{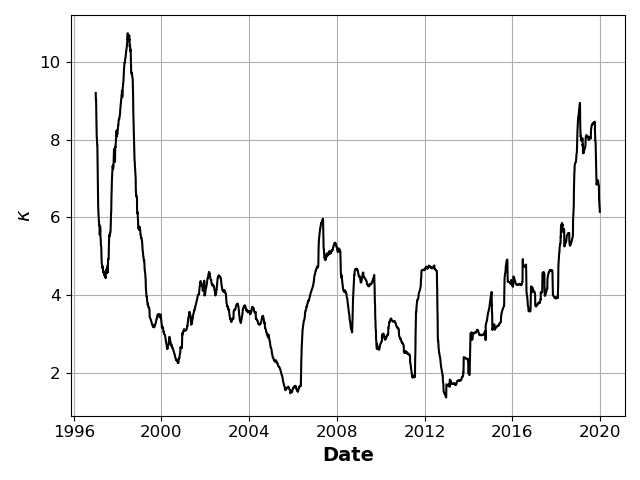}}
\subfigure[$\rho$]{\label{fig:a}\includegraphics[width=60mm]{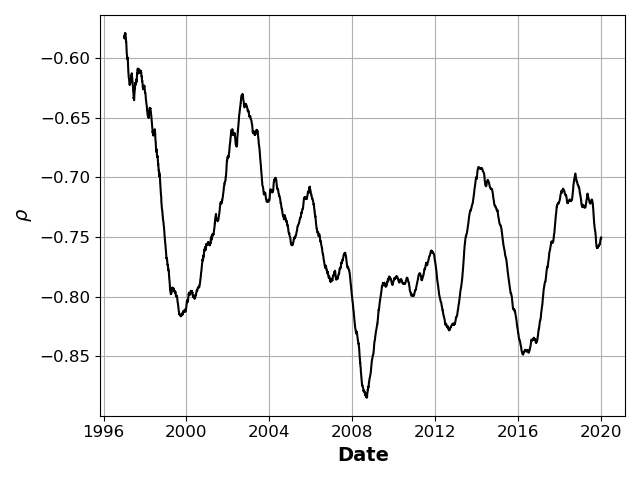}}
\subfigure[$\sigma$]{\label{fig:a}\includegraphics[width=60mm]{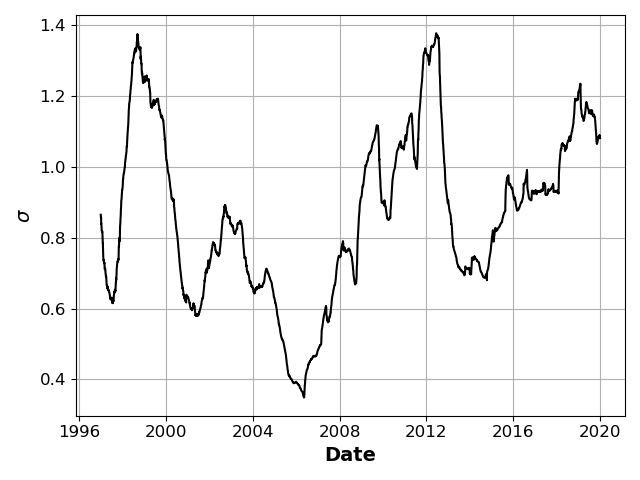}}
\subfigure[$\theta$]{\label{fig:a}\includegraphics[width=60mm]{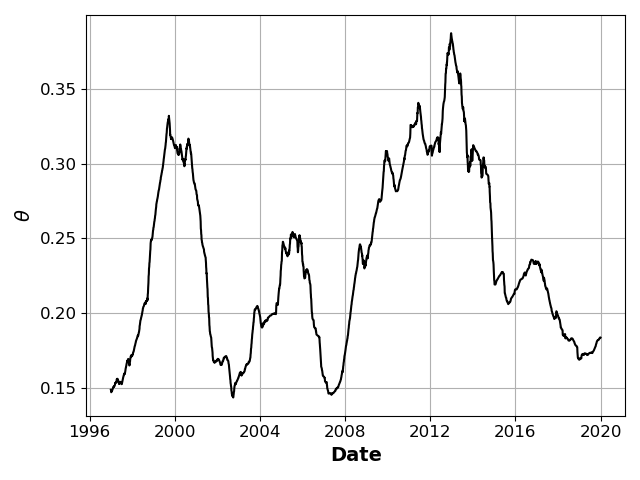}}
\subfigure[$\lambda$]{\label{fig:a}\includegraphics[width=60mm]{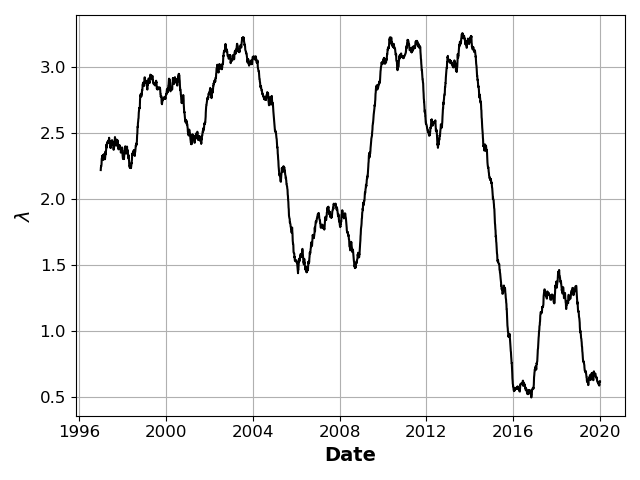}}
\subfigure[$p$]{\label{fig:a}\includegraphics[width=60mm]{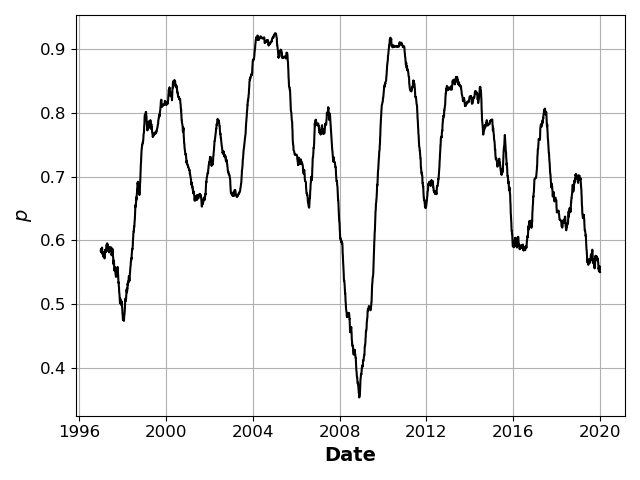}}
\subfigure[$\nu_1$]{\label{fig:a}\includegraphics[width=60mm]{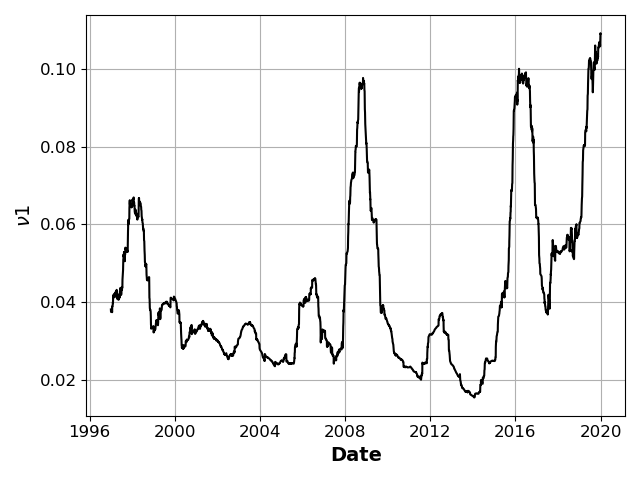}}
\subfigure[$\nu_2$]{\label{fig:a}\includegraphics[width=60mm]{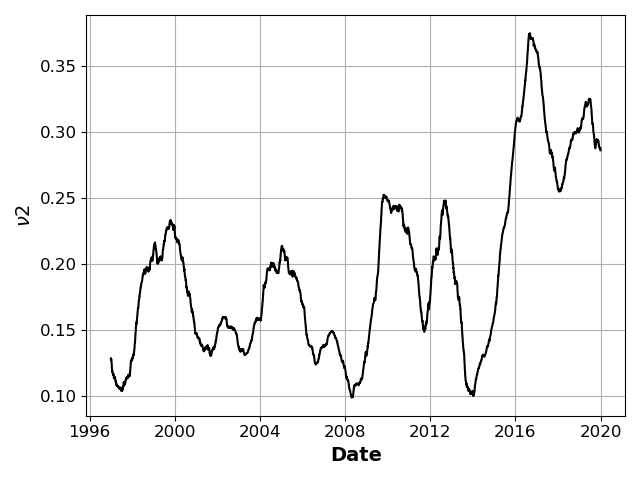}}
    \caption{
    The figures above show the daily estimation parameters of the BDJM. We smoothed the values with a rolling mean on the last 252 days. 
}
\label{fig:time_serie_params_BDJM}
\end{figure}

\clearpage

\section{Errors across strikes and maturities}
\label{sec:errors_across_strikes}

\begin{figure}[th!]
\centering     
\subfigure[$\tau=1$]{\label{fig:a}\includegraphics[width=50mm]{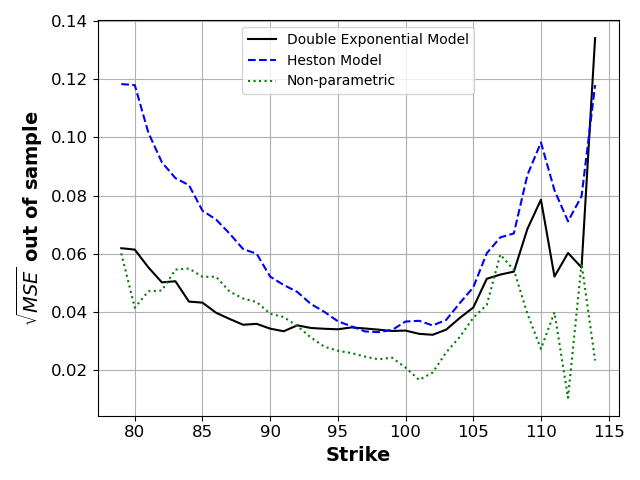}}
\subfigure[$\tau=5$]{\label{fig:b}\includegraphics[width=50mm]{fig/mse_os_L_5_Tcs7.png}}
\subfigure[$\tau=30$]{\label{fig:b}\includegraphics[width=50mm]{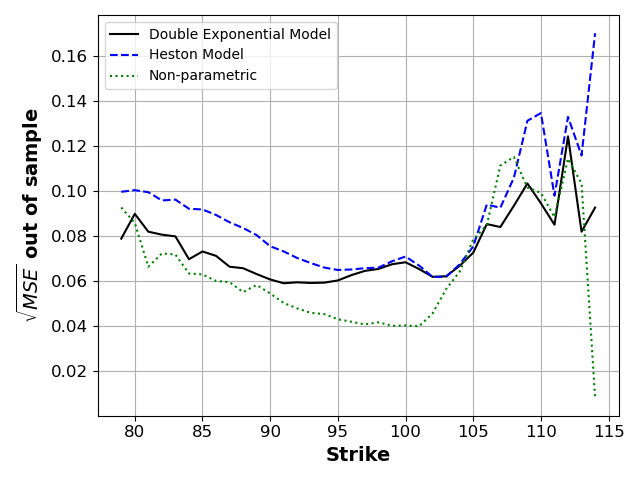}}
\caption{
The figures above show the same $\sqrt{MSE}$ per strike level as in figure \ref{fig:oos_cs_blind} computed on the subsample of options with a extremely-short time to maturity ($0<T\leq7$).
}
\label{fig:oos_time_serie_252}
\end{figure}

\begin{figure}[th!]
\centering     
\subfigure[$\tau=1$]{\label{fig:a}\includegraphics[width=50mm]{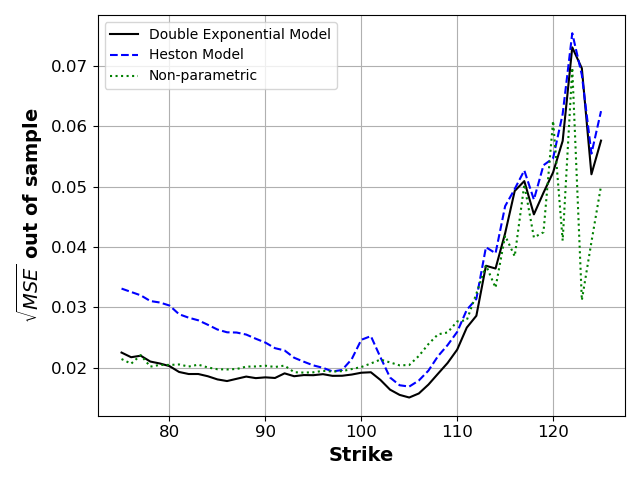}}
\subfigure[$\tau=5$]{\label{fig:b}\includegraphics[width=50mm]{fig/mse_os_L_5_Tcs30.png}}
\subfigure[$\tau=30$]{\label{fig:b}\includegraphics[width=50mm]{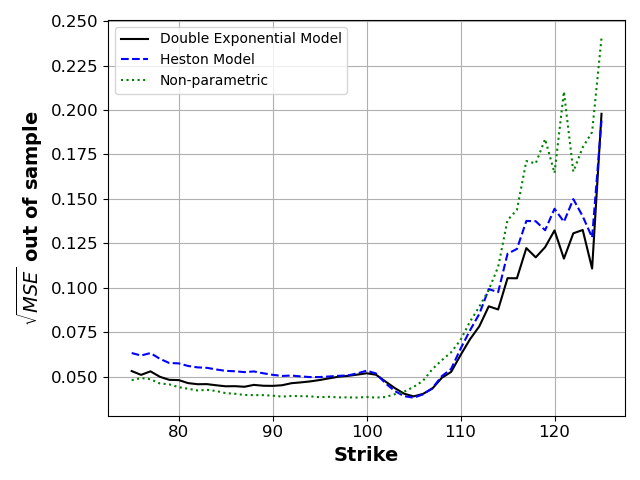}}
\caption{
The figures above show the same $\sqrt{MSE}$ per strike level as in figure \ref{fig:oos_cs_blind} computed on the subsample of options with a short time to maturity ($7<T\leq 30$).
}
\label{fig:oos_time_serie_252}
\end{figure}

\begin{figure}[th!]
\centering     
\subfigure[$\tau=1$]{\label{fig:a}\includegraphics[width=50mm]{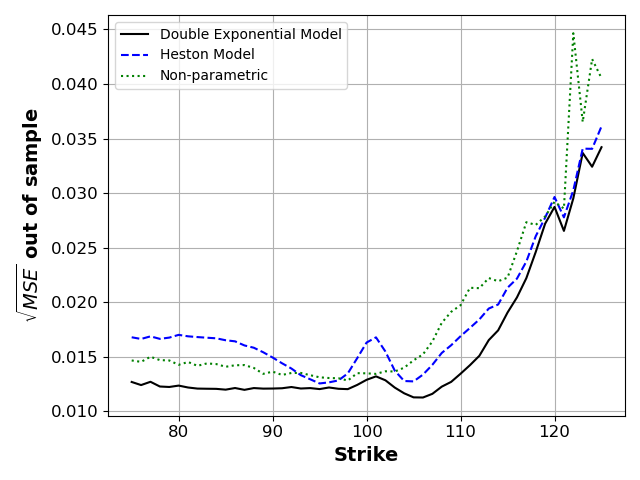}}
\subfigure[$\tau=5$]{\label{fig:b}\includegraphics[width=50mm]{fig/mse_os_L_5_Tcs90.png}}
\subfigure[$\tau=30$]{\label{fig:b}\includegraphics[width=50mm]{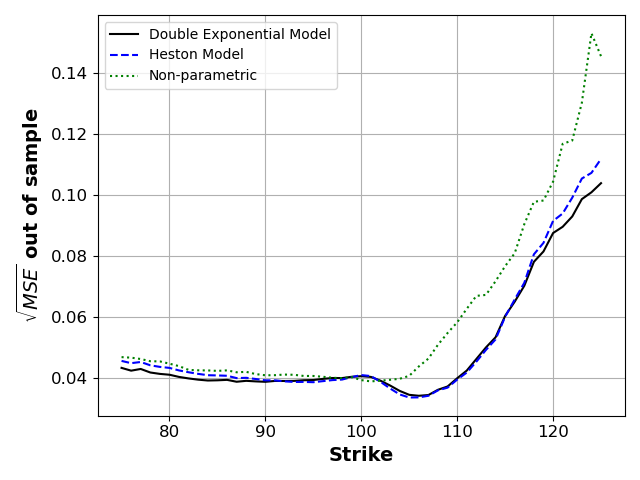}}
\caption{
The figures above show the same $\sqrt{MSE}$ per strike level as in figure \ref{fig:oos_cs_blind} computed on the subsample of options with a medium time to maturity ($30<T\leq 90$).
}
\label{fig:oos_time_serie_252}
\end{figure}

\begin{figure}[th!]
\centering     
\subfigure[$\tau=1$]{\label{fig:a}\includegraphics[width=50mm]{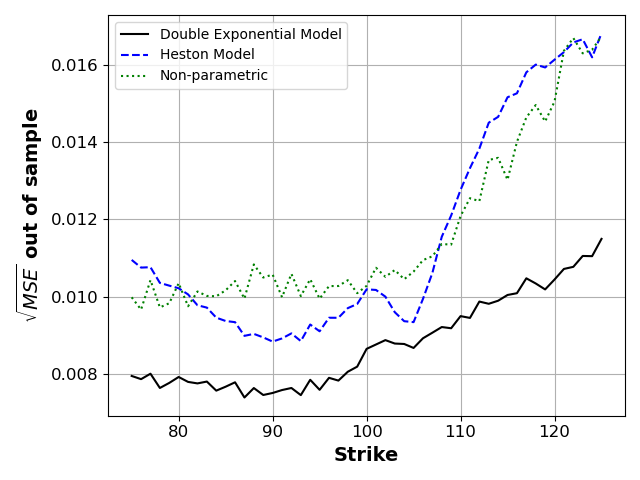}}
\subfigure[$\tau=5$]{\label{fig:b}\includegraphics[width=50mm]{fig/mse_os_L_5_Tcs500.png}}
\subfigure[$\tau=30$]{\label{fig:b}\includegraphics[width=50mm]{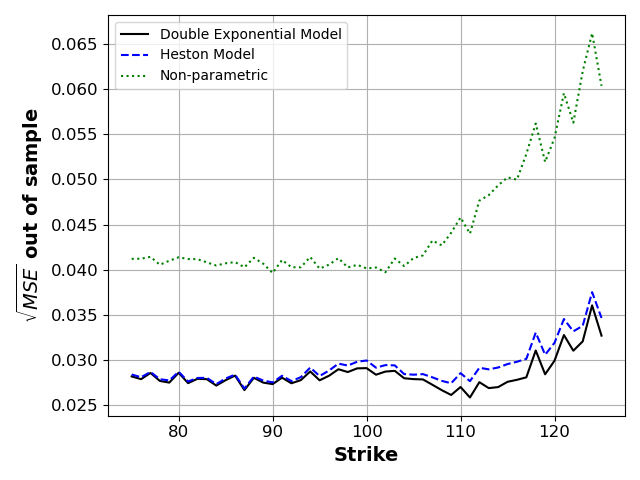}}
\caption{
The figures above show the same $\sqrt{MSE}$ per strike level as in figure \ref{fig:oos_cs_blind} computed on the subsample of options with a long time to maturity ($90<T\leq \infty $).
}
\label{fig:oos_time_serie_252}
\end{figure}

\clearpage


\section{Additional analysis}
\label{sec:additional_analysis}

\begin{table}[ht!]
  \centering
     \caption{This table expand on table \ref{table:replication_error} with additional percentiles of the replication error for each model. 
}
\begin{tabular}{lrrrr}
\toprule
{} &       BSM &        RF &        HM &      BDJM \\
\midrule
mean &   0.19380 &   0.20133 &   0.15501 &   0.14175 \\
std  &   0.42104 &   0.32173 &   0.39458 &   0.28051 \\
min  &   0.00000 &   0.00000 &   0.00000 &   0.00000 \\
5\%   &   0.00423 &   0.00754 &   0.00601 &   0.00595 \\
10\%  &   0.01083 &   0.01878 &   0.01225 &   0.01220 \\
15\%  &   0.01747 &   0.02963 &   0.01881 &   0.01879 \\
20\%  &   0.02453 &   0.04081 &   0.02570 &   0.02569 \\
25\%  &   0.03220 &   0.05255 &   0.03305 &   0.03299 \\
30\%  &   0.04073 &   0.06476 &   0.04085 &   0.04073 \\
35\%  &   0.05014 &   0.07787 &   0.04924 &   0.04908 \\
40\%  &   0.06080 &   0.09247 &   0.05831 &   0.05811 \\
45\%  &   0.07288 &   0.10846 &   0.06836 &   0.06807 \\
50\%  &   0.08686 &   0.12512 &   0.07958 &   0.07914 \\
55\%  &   0.10295 &   0.14439 &   0.09227 &   0.09162 \\
60\%  &   0.12183 &   0.16658 &   0.10681 &   0.10580 \\
65\%  &   0.14332 &   0.19171 &   0.12358 &   0.12206 \\
70\%  &   0.17039 &   0.22086 &   0.14330 &   0.14120 \\
75\%  &   0.20360 &   0.25554 &   0.16752 &   0.16450 \\
80\%  &   0.24989 &   0.29937 &   0.19887 &   0.19474 \\
85\%  &   0.31197 &   0.35633 &   0.24335 &   0.23784 \\
90\%  &   0.41471 &   0.43864 &   0.31658 &   0.30624 \\
95\%  &   0.65096 &   0.59113 &   0.48294 &   0.44869 \\
max  &  32.97002 &  30.43577 &  64.09164 &  29.19978 \\
\bottomrule
\end{tabular}

\label{table:replication_error_large}
\end{table}

\clearpage


\begin{onehalfspacing}   
\bibliographystyle{jf}
\bibliography{bib_econ}
\end{onehalfspacing}


\end{document}